\documentclass[12pt,a4paper,final]{iopart}

\usepackage{iopams}
\usepackage{aas_macros} 
\usepackage{graphicx}   
\usepackage{bm}         
\usepackage{amssymb}    
\usepackage{url}        
\usepackage{longtable}  
\usepackage{makecell}
\usepackage{mathrsfs}   

\newcommand{\sdata}{$Y_{\odot,r}$}                          
\newcommand{\nggn}{$(n,\gamma)\rightleftarrows(\gamma,n)$}  

\usepackage[breaklinks=true,colorlinks=true,linkcolor=blue,urlcolor=blue,citecolor=blue]{hyperref}

\begin{document}

\hspace{5.2in} \mbox{LA-UR-16-27225}

\title[Reverse engineering nuclear properties from rare earth abundances in the $r$ process]{Reverse engineering nuclear properties from rare earth abundances in the $r$ process}

\author{M.~R. Mumpower}
\address{Theoretical Division, Los Alamos National Lab, Los Alamos, NM 87545, USA}
\author{G.~C. McLaughlin}
\address{Department of Physics, North Carolina State University, Raleigh, NC 27695, USA}
\author{R. Surman}
\address{Department of Physics, University of Notre Dame, Notre Dame, IN 46556 USA}
\author{A.~W. Steiner}
\address{Department of Physics and Astronomy, University of Tennessee, Knoxville, TN 37996, USA}
\address{Physics Division, Oak Ridge National Laboratory, Oak Ridge, TN 37831, USA}

\date{\today}

\begin{abstract}
The bulk of the rare earth elements are believed to be synthesized in the rapid neutron capture process or $r$ process of nucleosynthesis. 
The solar $r$-process residuals show a small peak in the rare earths around $A\sim 160$, which is proposed to be formed dynamically during the end phase of the $r$ process by a pileup of material. 
This abundance feature is of particular importance as it is sensitive to both the nuclear physics inputs and the astrophysical conditions of the main $r$ process. 
We explore the formation of the rare earth peak from the perspective of an inverse problem, using Monte Carlo studies of nuclear masses to investigate the unknown nuclear properties required to best match rare earth abundance sector of the solar isotopic residuals. 
When nuclear masses are changed, we recalculate the relevant $\beta$-decay properties and neutron capture rates in the rare earth region. 
The feedback provided by this observational constraint allows for the reverse engineering of nuclear properties far from stability where no experimental information exists. 
We investigate a range of astrophysical conditions with this method and show how these lead to different predictions in the nuclear properties influential to the formation of the rare earth peak. 
We conclude that targeted experimental campaigns in this region will help to resolve the type of conditions responsible for the production of the rare earth nuclei, and will provide new insights into the longstanding problem of the astrophysical site(s) of the $r$ process.
\end{abstract}

\maketitle

\section{Introduction}\label{sec:intro}
One of the most intriguing open problems in nuclear astrophysics is the astrophysical site or sites of the production of the heaviest elements in the rapid neutron capture process, or $r$ process, of nucleosynthesis \cite{NRC03, NRC13}. 
The final elemental and isotopic abundances of the nuclei produced in the $r$ process can be seen in stars and found in meteorites \cite{Marti+85}. 
From these observations one tries to determine the astrophysical conditions under which the $r$ process occurs. 
Complicating this endeavor is a dearth of measurements of the properties of nuclei that participate in the $r$ process. 
The study of the $r$ process is therefore inherently an inverse problem - the output is known and the input must be determined. 

The output, the observed isotopic and elemental abundance patterns, e.g. \cite{Arlandini+99}, show a number of interesting features. 
There exists evidence of both a `weak' component to the $r$ process \cite{Qian+01, Travaglio+04, Aoki+05, Montes+07}, which produces material up until the region of atomic mass number $A\sim120$, and a `main' component which produces the rest of the heavier elements, $A\gtrsim120$ \cite{Wasserburg+96, Qian+07, Shibagaki+16}. 
A distinguishing factor between these two components is the scatter found in the elemental patterns of the weak component \cite{Sneden+08}, suggesting either variable conditions within a single type of astrophysical event or contributions from multiple sites. 
Here we focus on the main component, which is characterized by the robustly-produced heavier $r$-process peaks found at  $A\sim160$  and $A\sim195$, and likely also $A\sim130$ \cite{Roederer+12, Roederer+14}. 

One component of the input, the astrophysical conditions, must be such that there is a high neutron flux \cite{Burbidge+57, Cameron+57}. 
However, the precise amount of neutron-richness has not been established and neither has the degree of heating, or the range of outflow timescale, temperature or density. 
This has lead to a number of suggestions for the main $r$ process site, which include the traditional core collapse supernova and the merging of compact objects; see \cite{Arnould+07, Thielemann+11, Mumpower+16r} and references therein. 
Proposed $r$ process sites show marked differences in the evolution of the last stage of the $r$ process when nuclei slow their capture of neutrons and begin to decay back to stability, a phase known as `freeze-out'. 
Though many variations are possible, conditions during the final of the $r$ process can be generally classified as `hot' or `cold'. 
A hot $r$ process evolution goes through an extended equilibrium between neutron captures and its inverse reaction photodissociation, often written, \nggn. 
The freeze-out from equilibrium and the decay back to stability are triggered by an exhaustion of free neutrons. 
A cold $r$ process \cite{Wanajo+07} evolution has a short or non-existent \nggn \ phase where equilibrium fails due to the drop in temperature, followed by competition between neutron captures and $\beta$-decays. 

The other component of the input for the $r$ process is the yet to be measured nuclear properties of unstable neutron-rich nuclei. 
Theoretical nuclear models used in $r$-process calculations are well constrained and are mostly in agreement where data exists, however, the model predictions diverge as one approaches the driplines \cite{Kortelainen+12,Erler+13,Mumpower+15a}. 
Where there is disagreement between models, there is no experimental data and the majority of nuclei that have substantial impact on the final $r$-process abundances are in this category, see Refs.~\cite{Mumpower+15b, Liddick+16, Martin+16} for recent examples. 
The most important nuclear physics inputs for the $r$ process are masses, $\beta$-decays and neutron capture rates near closed neutron shells and in the rare earth region \cite{Mumpower+16r}. 

To solve an inverse problem, it is helpful to have the output as well determined as possible. 
The solar isotopic $r$-process abundances are defined by a residual procedure from the well constrained $s$-process abundances \cite{Kappeler+11}.  
In particular, the abundances of the rare earth elements, the peak at $A\sim160$, are some of the most precisely known in the solar system and in very metal-poor stars \cite{Lodders+09}. 
Further, the $r$-process rare earths are expected to be produced only in the main $r$ process, i.e. they don't have a weak component \cite{Sneden+08}. 
Therefore, the rare earth elements and the associated peak is an ideal choice for exploration of an $r$ process inversion technique \cite{Mumpower+16a}. 
 
Two distinct mechanisms have been previously proposed to explain rare earth peak formation. 
The first mechanism is dynamic formation of the peak during freeze-out \cite{Surman+97, Mumpower+12a}. 
In this scenario, material becomes hung up in the rare earth region during the decay back to stability. 
This mechanism requires a nuclear physics feature in this region responsible for the hangup and is sensitive to the late-time evolution of astrophysical conditions. 
The second mechanism is the formation of the peak by the deposition of fission fragments \cite{Schramm+71}. 
This possibility requires both multiple fission cycles from higher to lower atomic mass number and precisely tuned fission fragment distributions \cite{Goriely+13}. 
While a less aesthetically pleasing solution, it is also possible that the rare earth peak is formed by a combination of the two mechanisms. 
Experimental campaigns to produce the appropriate neutron-rich heavy isotopes and study their fission fragments are not possible now or in the foreseeable future. 
However, measurements of nuclei that are populated during the decay back to stability are possible in some cases currently and for others in near future. 
Ergo, the most sensible path forward is to try to confirm or eliminate the purely dynamical mechanism. 

In this manuscript, we take the observed rare earth abundance pattern and, for different types of astrophysical conditions, invert this abundance pattern to determine nuclear properties. 
We use common Bayesian inference techniques to find the region of the $NZ$-plane which dictates the shape and location of rare earth abundance pattern. 
The feedback provided by the observed rare earth abundances allows us to \textit{reverse engineer} the required trends in the nuclear masses that are responsible for the production of the rare earth peak. 
The larger strategy is to compare predictions of this type with future measurements, moving us closer toward an understanding of the astrophysical site of the main $r$ process. 
In section \ref{sec:method} we introduce this methodology and discuss all of the assumptions that go into our calculations. The propagation of nuclear model input changes is also covered in detail. 
In section \ref{sec:results} we give the results of these calculations and report the most favorable mass surface trends for each type of astrophysical conditions. 
In section \ref{sec:summary} we summarize. 

\section{Methodology}\label{sec:method}
The dynamical mechanism of rare earth peak formation requires a feature in the nuclear properties of rare-earth nuclei far from stability that causes a pileup of material as the $r$-process path moves toward stability. 
Pile-ups also occur in the main peaks which are thought to stem from  closed shell structure at neutron numbers, $N=82$ and $N=126$ extending far from stability. 
The rare earth peak is smaller than the main peaks and therefore a priori one does not know if it originates from a nuclear structure feature that has a sharp transition in N, but has a large extent in Z, similar to what might happen at subshell closure or a feature with a more gentle slope, but also more localized in the ($N$,$Z$) plane. 

It was suggested in the first work on the dynamical mechanism \cite{Surman+97} that the feature  originates from  a deformation maximum that leads to a `kink' in the neutron separation energies in this region. 
This kink was not the sharp feature seen at the closed shells, but instead of the smaller, more gentle type. 
In Ref.\ \cite{Surman+97} it was suggested that when this more localized feature is encountered during the late stages of the $r$ process, when \nggn \ equilibrium is freezing out, material funnels into the kink region to create a peak. 
Later it was shown that a similar feature in neutron capture rates can create a peak in conditions where the bulk of the $r$ process occurs out of equilibrium \cite{Mumpower+12a}. 
While the structure of the mass surface is clearly essential to understanding the formation of the rare earth peak in the dynamical mechanism \cite{Mumpower+12c}, the precise form of the structure has not yet been determined. 
The the rare earth peak height and location are sensitive to both the astrophysical conditions at late times in the $r$ process that govern the decay back to stability and the size and placement of the nuclear physics feature responsible for the pileup, which opens the possibility of constraining astrophysical conditions from the nuclear physics of the rare earth peak \cite{Mumpower+12b}. 

In this section we describe our methodology for treating the formation of the rare earths as an inverse problem. 
We utilize the sets of realistic astrophysical conditions described in \ref{method:astro} and in the context of our model uncover the feature in the nuclear masses in the rare earth region that best reproduces the final isotopic abundance pattern for each set of conditions. 
We assume that the rare earth peak forms via the dynamical mechanism during the freeze-out phase of the $r$ process, and that the nuclear feature responsible exists in the mass surface.
We further assume the nuclear feature can be described by a function of neutron number, $N$, proton number, $Z$. 
We use the Metropolis algorithm to explore the parameter space of our functional form for the nuclear masses. 
At each step in the Monte Carlo, we propagate the change in nuclear masses to other relevant quantities of the $r$ process as in Ref.~\cite{Mumpower+15a, Mumpower+15b}. 
The feedback, i.e. success or failure of a particular step in the Monte Carlo, is given by the match of the $r$-process network output to the observed rare earth abundances. 
We call this procedure the \textit{reverse engineering} framework and now discuss the motivation for our approach followed by the details of the methodology. 

\begin{figure}
 \includegraphics[width=160mm]{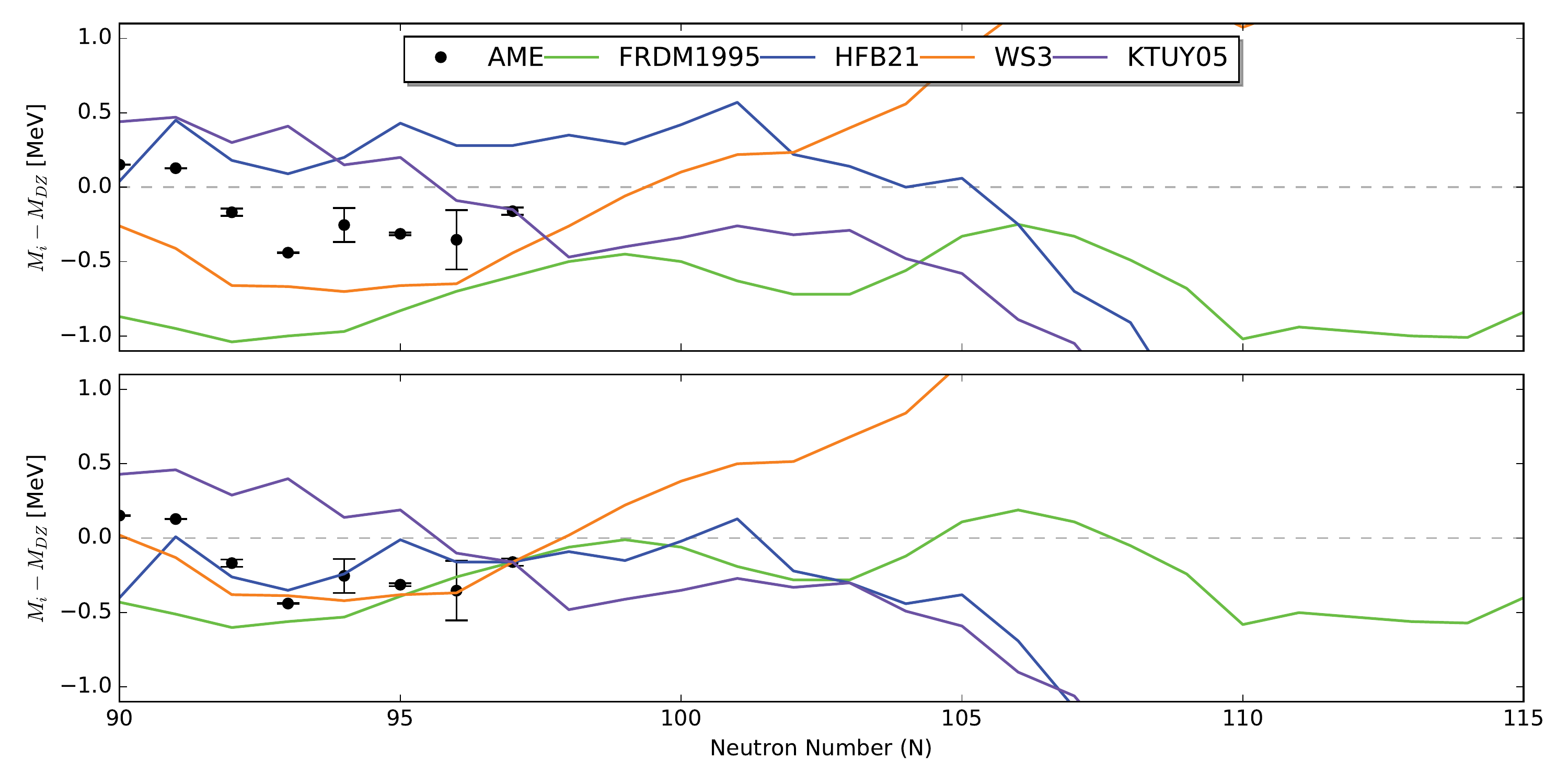}
 \caption{\label{fig:tmasses} Mass predictions from several mass models \cite{FRDM1995, HFB21, Liu+11, Koura+05} and measurements from the 2012 Atomic Mass Evaluation (AME2012) \cite{Audi+12} are compared to the Duflo-Zuker \cite{Duflo+95} masses for the $Z=60$ neodymium isotopic chain (top panel). Predicted trends in the mass surface are highlighted in the bottom panel, which shows the same mass comparisons, scaled to the last mass measurement at $N=97$. }
\end{figure}

\subsection{Mass surface parameterization}\label{method:mass}
\begin{figure}
 \includegraphics[width=160mm]{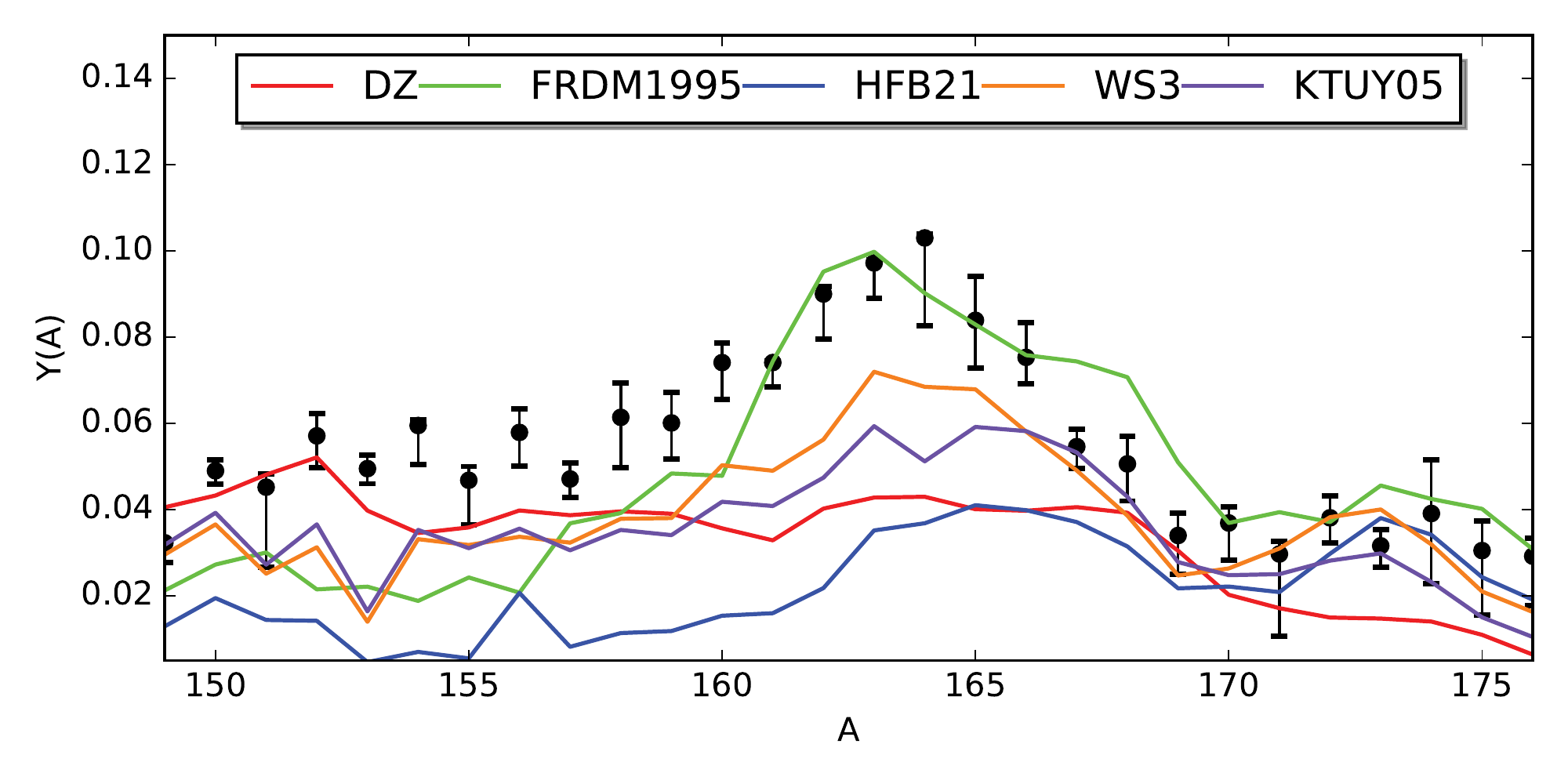}
 \caption{\label{fig:rep} The final rare earth abundances from several mass models shown in Fig.\ \ref{fig:tmasses} for hot $r$-process conditions - traj.\ 1 of this paper. In these simulations only the neutron separation energies change that go into the calculation of photodissociation rates. Unlike our reverse engineering framework, the $\beta$-decay rates from Ref.~\cite{Moller+03} and neutron capture rates Ref.~\cite{Rauscher+00} remain unchanged. Solar data from Ref.~\cite{Arnould+07}. }
\end{figure}

We focus on nuclear masses between the $N=82$ and $N=126$ closed shells, since in previous work is was shown that the most crucial masses for the dynamical mechanism are centered near $N\sim100$ \cite{Surman+97, Mumpower+12a, Mumpower+12b}. 
The mass predictions of several mass models commonly used in $r$-process calculations are compared in Fig.~\ref{fig:tmasses} for the neodymium isotopic chain. 
Note the overall predictions of the mass models roughly agree to $\pm0.5$ MeV where data is available and diverge at higher neutron numbers. 
The nuclear data important for dynamical rare earth peak formation does not, however, depend on the absolute values of the masses but mass differences, e.g. neutron separation energies and $Q$-values. 
Therefore, the general trends in the mass surface are of greatest importance for the formation of the rare earth peak. 
These predicted trends, compared in the bottom panel of Fig.~\ref{fig:tmasses}, show markedly different shapes and behaviors. 
When applied to $r$-process simulations, different mass models produce rare earth peaks with varying degrees of success \cite{Mumpower+12b}, as shown in Fig.~\ref{fig:rep}. 
Some theoretical nuclear mass models, such as the largely empirical Duflo-Zuker (DZ) \cite{Duflo+95} model, show no feature in this region, and thus $r$-process simulations run with these models do not exhibit dynamical rare earth peak formation, e.g. Fig.~1 from \cite{Mumpower+16a}. 
Others, such as the 1995 version of the FRDM \cite{FRDM1995}, find that the consequences of deformation are the prediction of a feature in the mass surface of sufficient depth to prompt rare earth peak formation, for certain ranges of astrophysical conditions. 

If the rare earth peak forms via the dynamical mechanism, the structure in the mass surface that is responsible for peak formation may be localized in $Z$ and $N$ or exhibit a more global, smooth trend in the region. 
We would like to use an algorithm that allows for both possibilities. 
One possible starting point is to begin with a baseline mass model, and then produce small changes to each nuclear mass prediction in the rare earth region via Monte Carlo sampling until the peak has been produced.
The problem with such an approach is the number of parameters quickly exceeds the number of observable rare earth abundances, as one parameter is needed per nucleus.
This leads to an overdetermined system, which in all likelihood would converge to solutions that are not physically meaningful. 

In our approach, we begin with the Duflo-Zuker (DZ) mass model \cite{Duflo+95}, a model that exhibits no regional trend in the rare earth mass surface, and hence no rare earth peak. 
To account for the variety in possible mass surfaces (local or global) we modify the baseline DZ masses with an extra term, 
\begin{equation}
M(Z,N) = M_{DZ}(Z,N) + a_N e^{-(Z-C)^2/2f}
\label{eq:masses}
\end{equation}
where $M(Z,N)$ is the new mass prediction for the nucleus with $Z$ protons and $N$ neutrons, $M_{DZ}(Z,N)$ is the baseline DZ mass, and the second term on the right hand side contains the mass modification parameters that will be run through the Monte Carlo procedure. 
Each isotone in the region, $N$ ranging from $95$ to $115$, is assigned a unique coefficient, ${a_N}$. 
For a given neutron number, $a_N$ controls the overall magnitude and sign of the mass change to the DZ model. 
The parameter $C$ controls the center of the strength of the mass changes in proton number. 
If $Z=C$, the exponential term goes to unity and the mass changes from the $a_N$ are maximal. 
We also incorporate a fall off parameter, $f$, which controls the rate at which the mass modifications return return to zero, and the total mass prediction returns to the baseline DZ predictions in proton number. 
The fall off parameter ensures that any feature found in the mass surface responsible for peak formation disappears closer to stability, in accordance with measurements. 
The absence of an imposed functional form for the behavior of the mass surface as a function of neutron number, $N$, allows the algorithm to freely determine whether a kink structure exists or not. 
In principle there is the danger of converging on unphysical solutions with $a_N$ allowed to flow freely, however, given the astrophysical conditions studied here, the solutions we find are always well behaved. 

\subsection{Neutron capture}\label{method:ncap}
The formation of the rare earth peak is also sensitive to the neutron capture rates in the region \cite{Mumpower+12c}. 
Neutron capture rates, which depend on the mass surface, in our reaction network are calculated using the Los Alamos statistical Hauser-Feshbach (HF) code, CoH (version 3.3.3) \cite{Kawano+06, Kawano+08, Kawano+10, Kawano+16a}. 
The most important model ingredients to this code besides the mass surface are the assumed level density (LD), the $\gamma$-strength function ($\gamma$SF) and the particle optical model potential (OMP) \cite{Beard+14}. 

\begin{figure}
 \includegraphics[width=160mm]{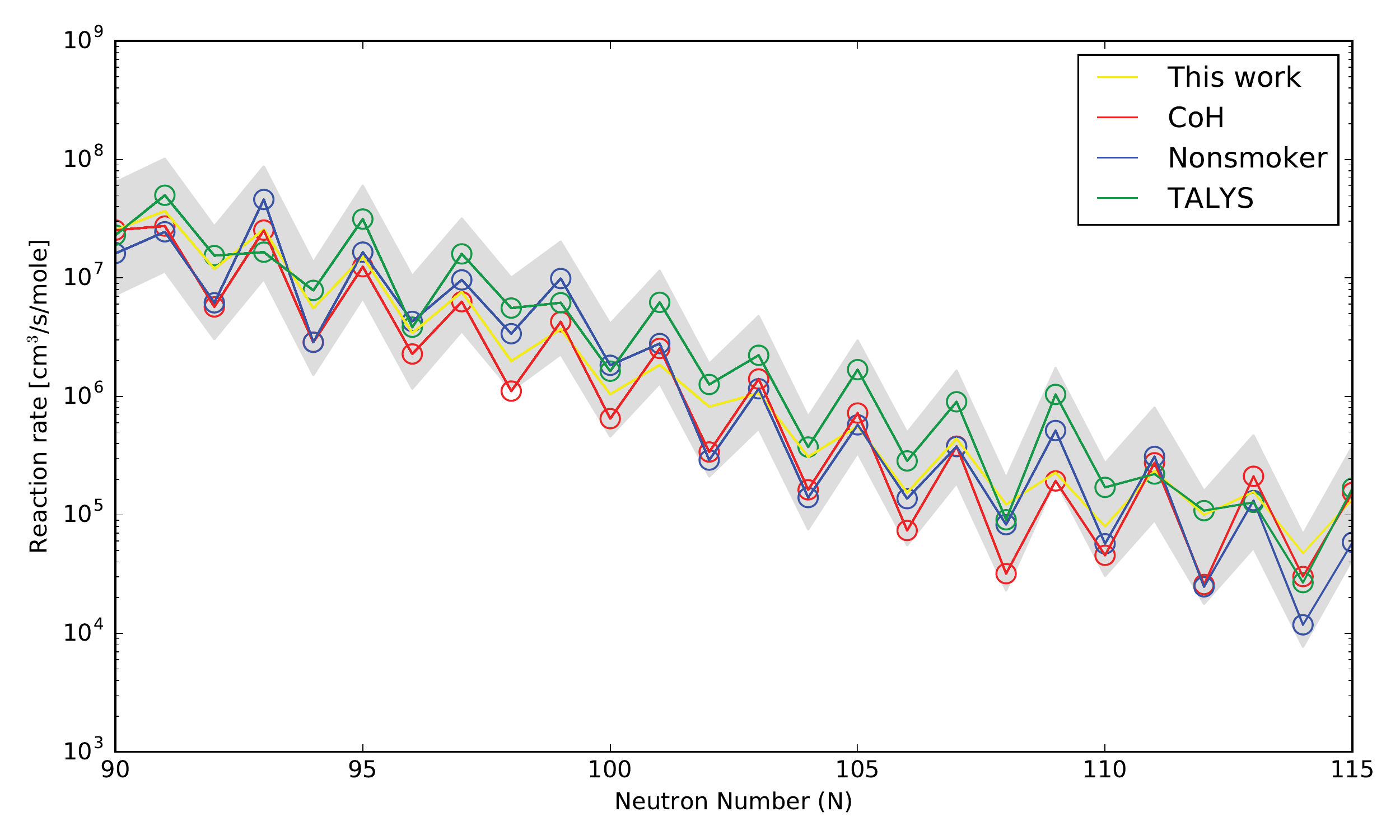}
 \caption{\label{fig:tncap} Theoretical astrophysical reaction rates along the $Z=60$ neodymium isotopic chain for several statistical model codes \cite{Kawano+16a,Rauscher+00,TALYS} and the baseline values used in this work evaluated at a temperature $T=1.0$ GK. Shaded band shows a factor of three variation from the average of the three statistical model predictions. }
\end{figure}

No neutron capture data exists for the neutron-rich nuclei that participate in the $r$ process, thus the model ingredients all depend on theoretical calculations. 
CoH uses the Gilbert-Cameron level density \cite{Gilbert+65} which is a hybrid description that uses a constant temperature model at low energies and matches to a Fermi gas model in the high energy regime that also includes shell corrections by Ignatyuk et al.~\cite{Ignatyuk+75}.
The $\gamma$-ray transmission coefficients are constructed using the generalized Lorentzian $\gamma$SF \cite{Kopecky+90} and the Koning-Delaroche global OMP is from Ref.~\cite{Koning+03}. 
Fig.~\ref{fig:tncap} shows the results of CoH calculations with DZ masses for the neodymium isotopic chain compared to neutron capture rate compilations commonly used in $r$-process simulations: NONSMOKER rates \cite{Rauscher+00} with FRDM masses and TALYS \cite{TALYS} rates with HFB masses. 
While in some regions of the nuclear chart these rate compilations can disagree by orders of magnitude, in the rare earth region they agree within about a factor of 3 with similar model inputs. 

The calculations of neutron capture rates are very time consuming, and so to good approximation one can capture the dependency of the change in masses by calculating \cite{Surman+97}
\begin{equation}
\lambda_{n,\gamma}(Z,N) = \exp \big[ a(N,T) + b(N,T) S_{n} + c(N,T) S_{n}^{2} \big]
\label{eq:ncap}
\end{equation}
where $a(N,T)$, $b(N,T)$ and $c(N,T)$ are temperature-dependent parameters for a given isotone, $N$, and $S_{n}$ is the one neutron separation energy. 
The units of $\lambda_{n,\gamma}$ are taken to be sec$^{-1}$. 
The rates are fit to the predictions of CoH using the baseline DZ masses on a temperature grid which fixes $a$, $b$ and $c$ for a given temperature $T$. 
These temperature-dependent parameters do not change throughout any of the Monte Carlo calculations. 
Using this approximation we find around a factor of 2 (or less) change in rate for $\Delta S_{n}=500$ keV, which is in agreement with typical factors obtained when performing the entire neutron capture rate calculation over again using CoH \cite{Mumpower+15b} or TALYS \cite{Mumpower+15a}. 
The results of this approximation are compared to the various neutron capture rate compilations in Fig.~\ref{fig:tncap}. 
As seen from Fig.~\ref{fig:tncap} the formulation is a reasonably good approximation to neutron capture rates and their dependence on nuclear masses. 
Additional uncertainties in neutron capture rates that stem from the $\gamma$SF are discussed in Sec.~\ref{sec:uncert}. 

\subsection{$\beta$-decay}\label{method:beta}
\begin{figure}
 \includegraphics[width=160mm]{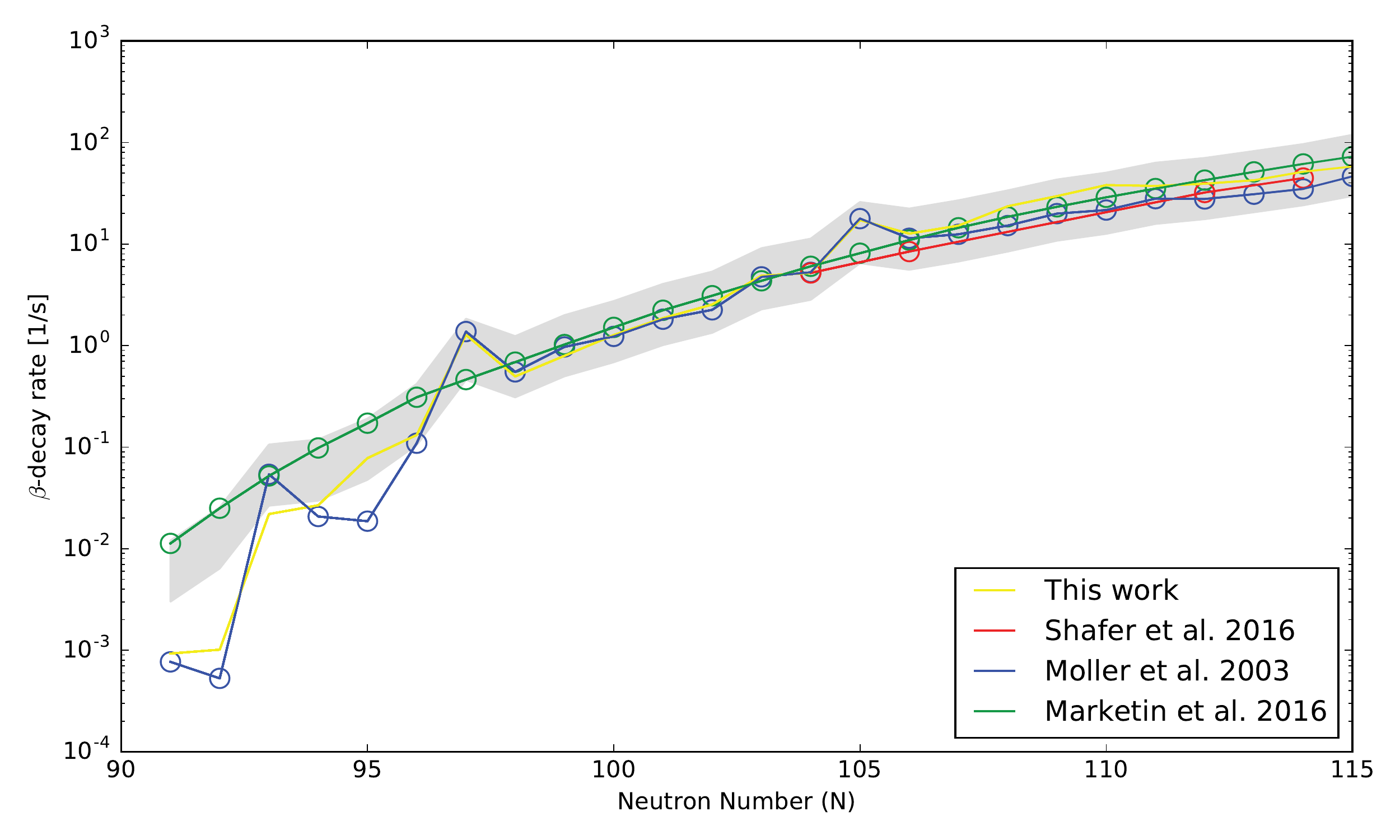}
 \caption{\label{fig:tbeta} Theoretical $\beta$-decay rates for the $Z=60$ neodymium isotopic chain from two compilations (Moller et al.~\cite{Moller+03} and Marketin et al.~\cite{Marketin+16}) along with recent predictions from Shafer et al.~\cite{Shafer+16} and the baseline values predicted by this work. }
\end{figure}

The $\beta$-decay properties of interest for $r$-process nucleosynthesis and rare earth peak formation are half-lives and delayed neutron emission probabilities \cite{Surman+97, Mumpower+12b, Lorusso+15, Shafer+16, Wu+16, Mumpower+16r}. 
Both quantities depend on a theoretical description of the $\beta$-strength function, $S_\beta$, as well as the assumed mass surface of neutron-rich nuclei \cite{Moller+03}. 
Close to stability, the energy window for $\beta$-decay, $Q_\beta$, is small, and the theoretical calculations are most sensitive to the details of the predicted nuclear structure. 
Further from stability, model predictions of $\beta$-decay rates are more consistent, with some variation coming from the assumed mass surface. 
In general, theoretical models of $\beta$-decay rates in the $N>95$ region important for rare earth peak formation vary by only a factor of two or so, as shown for the neodymium isotopic chain in Fig.~\ref{fig:tbeta}. 
Recently, it was shown that different Skyrme interactions in the Finite-Amplitude Method \cite{Shafer+16} agree quite closely with the results of older Quasi-particle Random Phase Approximation (QRPA) calculations \cite{Moller+03}. 
Variations of rates within the grey band of Fig.~\ref{fig:tbeta} may impact rare earth peak formation as shown in Ref.~\cite{Shafer+16}. 

To study the variation in $\beta$-decay rates on the order of the size of the grey band in Fig.~\ref{fig:tbeta} we propagate the Monte Carlo mass surface changes from Sec.~\ref{method:mass} to the half-lives by recalculating: 
\begin{equation}
T^{-1}_{1/2} = \sum_{0\leq E_i\leq Q_\beta} S_\beta(E_i) \mathscr{F}(Z,R,Q_\beta - E_i)
\label{eq:T12}
\end{equation}
where $S_\beta$ is the $\beta$-strength function evaluated at excitation energy $E_i$ in the daughter nucleus and $\mathscr{F}$ is the Fermi function evaluated with proton number $Z$, nuclear radius $R$ and energy window $Q_\beta-E_i$. 
The summation runs over all the Gamow-Teller strength from the Quasi-particle Random Phase Approximation (QRPA) solutions of Ref.~\cite{Moller+97}. 
A change to nuclear masses can modify $S_\beta$, $\mathscr{F}$, and the limits of the summation in Eq.\ (\ref{eq:T12}). 
The majority of the mass dependence sits in the phase space piece $\mathscr{F}$, which goes as the fifth power of the energy for allowed decays. 
The computationally-expensive nuclear matrix elements, in contrast, depend much less strongly on the masses. 
We therefore explore the impact of the change in $Q_\beta$ on $\beta$-decay properties, which impacts both $\mathscr{F}$ and the summation limits, while leaving the $\beta$-strength function unchanged from Ref.~\cite{Moller+97}. 
This approximation has been used previously in Refs.~\cite{Mumpower+15a, Mumpower+15b}. 
We find differences in $Q_\beta$ of $500$ keV yield roughly up to a factor of $2$ or so change in the half-life. 

We also propagate the mass changes to the prediction of $\beta$-delayed neutron emission probabilities, which we calculate by using the recently pioneered coupled QRPA+HF method \cite{Mumpower+16b}.
In this method, neutron-gamma competition is tracked through subsequent generations during the statistical decay until all available excitation energy is spent.
The probability to emit $j$-neutrons is given as a recursive convolution of level populations,
\begin{equation}
P_{jn}(E_\textrm{gs}) = \sum_{i=0}^{k-1} \mathscr{P}_{j}(E_i) p_{j}(E_i, E_\textrm{gs})
                      + \sum_{{k^\prime}=0}^{k-1} \mathscr{P}_{j-1}(E_{k^\prime}) q_{j-1}(E_{k^\prime}, E_\textrm{gs})
\label{eq:Pjn}
\end{equation}
where $\mathscr{P}_j$ indicates the level population for the $j$-th compound nucleus, $p_{j}$ gives the probability to emit a $\gamma$-ray from an excited state to the ground state in the $j$-th compound nucleus, $q_{j-1}$ represents the probability to emit a neutron from the previous compound nucleus to the $j$-th, the summations run over all levels which may feed the compound state $k$ in $j$-th compound nucleus, and the initial level population is given by the $\beta$-decay strength function, $\mathscr{P}_0(E_k)=S_\beta(E_k)$. 
For consistency, the same strength data as in the half-life calculation is used in the initial population of the compound daughter nucleus. 
Details of the QRPA+HF method and further discussion of Eq.\ (\ref{eq:Pjn}), including the definition of the $p$ and $q$ functions, are given in Ref.\ \cite{Mumpower+16b}. 

\begin{figure}
 \includegraphics[width=160mm]{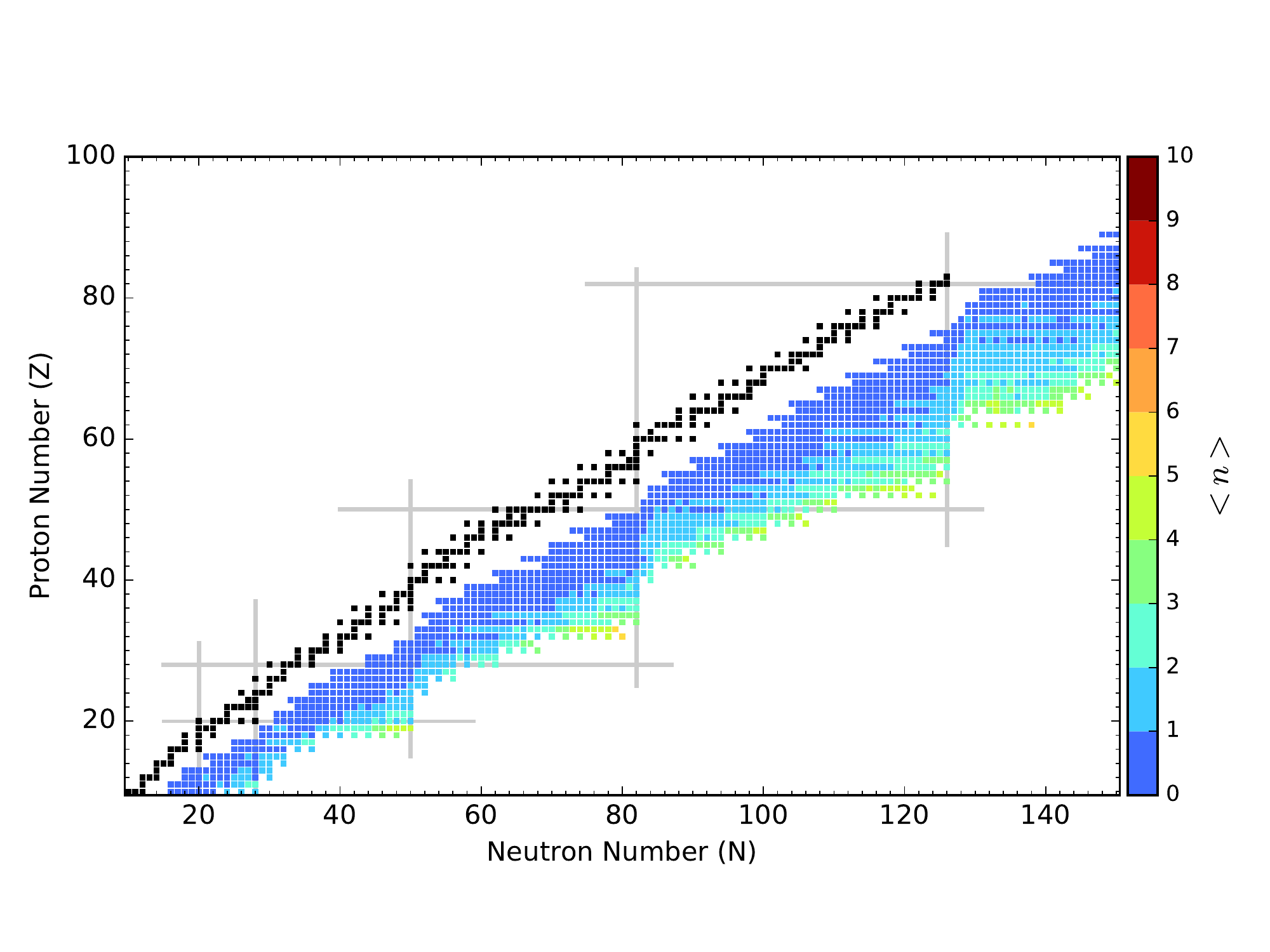}
 \caption{\label{fig:avgN} Average number of neutrons emitted after $\beta$-decay using the QRPA+HF model assuming Duflo-Zuker masses. }
\end{figure}

Both $Q_\beta$ and the neutron separation energies, $S_n$, are input arguments for the $p$ and $q$ functions, and a clear dependency of $P_{jn}$ on these quantities is masked by the required convolution in Eq.\ (\ref{eq:Pjn}).
Thus it is important to propagate the mass changes all the way through in Eq.\ (\ref{eq:Pjn}), as we do here.  
While, the intricate nature of the equation makes it difficult to gauge how mass changes will impact $P_{jn}$ values, we do note that using DZ masses, a clear trend emerges with on average less neutrons emitted after $\beta$-decay than in the case of another popular mass model FRDM1995. 
This can be seen by comparing the average number of neutrons emitted after $\beta$-decay, $\langle n \rangle$, in Fig.~\ref{fig:avgN} to the results displayed in Fig.\ 6 of Ref.~\cite{Mumpower+16b}. 

\subsection{Fission}
Because we seek a solution for rare earth peak formation that depends on the dynamical mechanism we employ a simple treatment of fission. 
We assume the dominant fission mechanism is the spontaneous fission of nuclei with $A>250$, and we take the fission daughter product distributions to be a 57/43 split. 
This ensures that fission fragments fall in the $A\sim 130$ peak region and do not directly influence rare earth peak formation. 
In the future our framework can be expanded to models of fission that include the deposition of fission daughter products on the rare earth peak region. 
Fission inputs for $r$-process calculations remains an exciting and active area of research \cite{Eichler+15, Moller+15}. 

\subsection{Network calculations}
The nuclear physics ingredients discussed above are coupled to astrophysical trajectories in a nuclear reaction network code. 
For our calculations we use the fast $r$-process network which is optimized to explore the freeze-out phase, most recently used in Ref.~\cite{Mumpower+16r}. 
This network has channels for neutron capture, photodissociation, $\beta$-decay, $\beta$-delayed neutron emission, and fission and has robust support for modification of nuclear physics inputs. 
Further discussion of this $r$-process network can be found in Refs.~\cite{Mumpower+12a, Mumpower+12b, Mumpower+12c}. 

\subsection{Astrophysical Conditions}\label{method:astro}
\begin{figure}
 \includegraphics[width=160mm]{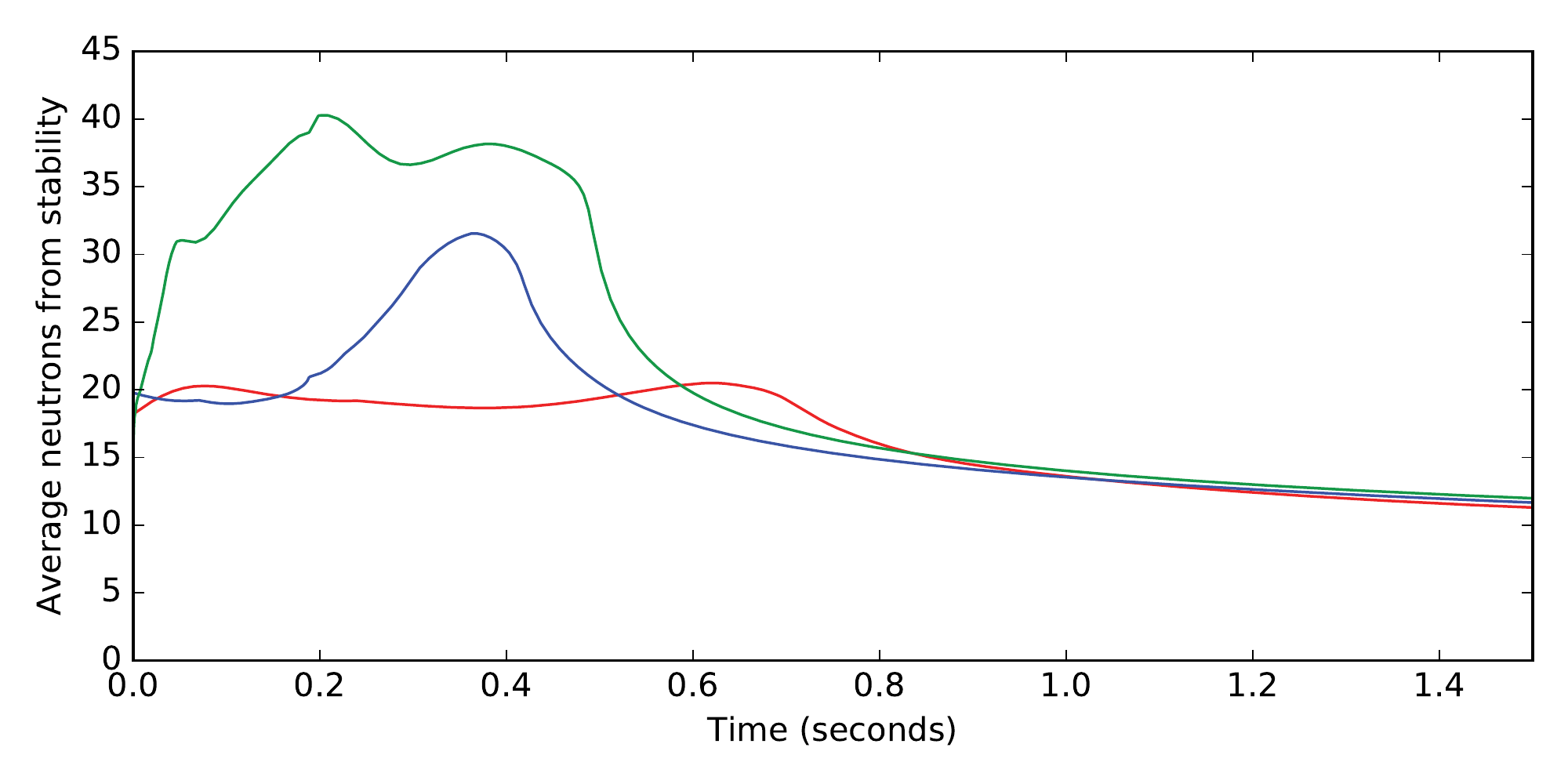}
 \caption{\label{fig:dN} Average number of neutrons from stability for hot - traj.\ 1 (red), cold - traj.\ 4 (blue) and very neutron-rich cold - traj.\ 8 (green) astrophysical conditions during the freeze-out phase of the $r$ process. Rare earth peak formation occurs when the nuclei begin to decay back to stability in each case. A time of zero indicates when the temperature has reached $T=2.0$ GK. }
\end{figure}

Since the astrophysical site of the $r$ process is uncertain, we select astrophysical trajectories that cover a broad spectrum of possibilities. 
To model a hot $r$-process which goes through an extended duration \nggn \ equilibrium phase we select trajectories from parameterized winds entropies $30$, $200$, and $110$ in units of $k_{B}$/baryon with timescales $\tau=70$, $80$, and $160$ in units of ms and electron fractions $Y_e=0.2$, $0.3$, and $0.2$, respectively \cite{Mumpower+12a}. 
We label these trajectories as trajectories 1, 2, and 3, respectively. 
For moderately cold $r$-process components with a short duration \nggn \ equilibrium we choose a trajectory from a detailed supernova model with reheating \cite{Arcones+07}, a wind parameterized as in \cite{Panov+09} with entropy of $75$ in units of $k_B$/baryon, $\tau=75$ ms and $Y_e=0.2$, and an extreme trajectory with very fast evolution, parameterized as in \cite{Mumpower+12b}: entropy of $125$ in units of $k_{B}$/baryon, initial timescale of $\tau=80$ ms, $Y_e=0.2$, and freeze-out power law of $n=6$. 
These are labeled trajectories 4, 5, and 6, respectively. 
For very neutron-rich cold $r$-process components we use trajectories, labeled 7, 8, and 9, from simulations of Refs.~\cite{Goriely+11, Just+15}. 
A distinction between these three astrophysical evolutions is shown in Fig.~\ref{fig:dN}. 
The very neutron-rich cold trajectories have an $r$-process path which ventures closer towards the neutron dripline relative to the hot and cold trajectories. 

We summarize the choice of astrophysical trajectories in Table \ref{table1}. 
The high entropy conditions are possible in neutrino-driven wind environments, however we employ lower electron fractions than typically found in detailed supernova models \cite{Martinez-Pinedo+14} without exotic physics \cite{McLaughlin+99}. 
The low entropy conditions may be possible in an accretion disk wind with the expected values of $Y_{e}$. 
The very neutron-rich cold conditions used are possible in the tidal tail ejecta from neutron star-neutron star or neutron star-black hole mergers. 

\begin{center}
\begin{longtable}{@{\extracolsep{\fill}} c|c|c}
\caption{\label{table1} List of astrophysical trajectories used in this work. }
\setlength{\tabcolsep}{2cm}\\
Trajectory & Type & Description\\
\hline\hline
\endhead
traj.\ 1 & hot              & Low entropy wind with low $Y_e$ \cite{Mumpower+12a} \\
traj.\ 2 & hot              & High entropy wind with moderate $Y_e$  \cite{Mumpower+12a} \\
traj.\ 3 & hot              & High entropy wind with low $Y_e$ \cite{Mumpower+12a} \\
traj.\ 4 & cold             & Detailed supernova model with reheating \cite{Arcones+07} \\
traj.\ 5 & cold             & Cold wind parameterization \cite{Panov+09} \\
traj.\ 6 & cold             & Very fast cold outflow parameterization \cite{Mumpower+12b} \\
traj.\ 7 & very n-rich cold & Outflow from merger simulation \cite{Goriely+11} \\
traj.\ 8 & very n-rich cold & Outflow from merger simulation \cite{Just+15} \\
traj.\ 9 & very n-rich cold & Outflow from merger simulation  \cite{Goriely+11}
\end{longtable}
\end{center}

\subsection{Algorithm}
For each of our chosen astrophysical trajectories, we determine the mass surface that best reproduces the solar rare earth peak using our reverse engineering framework, based on the widely-used Metropolis algorithm \cite{Metropolis+53}. 
The Bayesian approach we employ here is based on the popular techniques that are applicable to a wide variety of problems in science \cite{Brooks+03, vonToussaint+11}. 

Our Monte Carlo parameters are defined by those appearing in the second term of Eq.\ (\ref{eq:masses}): the $a_N$'s, $C$, and sometimes $f$ (if it is not held fixed).
Each of these parameters is varied independently using Gaussian distributions with width $25$ keV for the $a_N$'s, $0.1$ for $C$, and $0.5$ for $f$.
We start each Monte Carlo run by setting these parameters to zero, so that we begin with the baseline DZ prediction (no rare earth peak).

At the beginning of each Monte Carlo step we vary all the masses that enter into the reaction network by computing new parameter values and applying Eq.\ (\ref{eq:masses}). 
For the DZ mass model this consists of roughly 500 nuclei that are in the range $A\sim150$ to $A\sim180$. 
Next, we propagate the changes produced by the small variations in the Monte Carlo parameters to the remainder of the $r$-process nuclear physics inputs as defined in Sec.~\ref{method:mass} to \ref{method:beta}. 
The $r$-process network is then run with the nuclear physics ingredients from this particular set of Monte Carlo parameters. 

The likelihood function for a given Monte Carlo step, $m$, is defined by
\begin{equation}
\mathcal{L}(m) = \exp \big[ -\chi^2_\textrm{r}(m)/2 - \chi^2_\textrm{m}(m)/2 \big]
\label{eq:L}
\end{equation}
where represents $\chi^2_{r}$ the chi-squared function for matching the network abundance output to the solar isotopic pattern and $\chi^2_{m}$ is the chi-squared function for matching the theoretical masses to the measured masses in the 2012 compilation \cite{Audi+12}. 

More specifically, the chi-squared function for the $r$ process for a given Monte Carlo step, $m$, is defined as
\begin{equation}
\chi^2_\textrm{r}(m) = \frac{1}{\Delta Y^{2}} \sum^{A=180}_{A=150} \big[Y_{\odot,r}(A) - Y(A)) \big]^{2}
\label{eq:chi2r}
\end{equation}
where $\Delta Y$ is the average observational uncertainty of the abundances in the rare earth region, \sdata$(A)$ \ is the isotopic solar $r$-process residual, $Y(A)$ is the isotopic sum of the output of our network calculation and the summation runs over $A$, the atomic mass number.
The summation is limited in extent because we are only focused on the production of a local abundance feature, the rare earth peak.
The lower limit in the solar isotopic residuals may not be defined, so for each $A$ we take an approximate value of the observational uncertainty in the abundances of the rare earth region as $\Delta Y=0.1$ dex. 

We define a similar chi-squared function for the masses for a given Monte Carlo step, $m$,
\begin{equation}
\chi^2_\textrm{m}(m) = \frac{1}{\Delta M^{2}} \sum_{Z,N} \big[M_{AME}(Z,N) - M(Z,N) \big]^{2}
\label{eq:chi2m}
\end{equation}
where $\Delta M\sim 400$ keV is taken to be the average root-mean-square value for DZ compared to the 2012 Atomic Mass Evaluation (AME2012), $M_{AME}(Z,N)$ is the measured mass and $M(Z,N)$ is defined in Eq.~(\ref{eq:masses}) and the summation runs over all the nuclei with measured values in the AME2012.

To gauge the success or failure of a Monte Carlo step we compute the acceptance ratio,
\begin{equation}
\alpha(m) = \frac{\mathcal{L}(m)}{\mathcal{L}(m-1)}
\label{eq:alpha}
\end{equation}
where $\mathcal{L}(m)$ is the likelihood function for the given step and $\mathcal{L}(m-1)$ is the likelihood function for the previous step.
The baseline calculation using only DZ masses with all other parameters set to zero defines the likelihood function of the first step, $\mathcal{L}(0)$.
If $\alpha(m)\geq 1$, the candidate step, $m$, is more likely than the previous, so we accept the step and update the parameters to the new values.
If $\alpha(m)< 1$, the candidate step is taken with probability $\alpha(m)$, otherwise the step is rejected and the parameters are reset to those defined by the last successful step.

\begin{figure}
 \includegraphics[width=160mm]{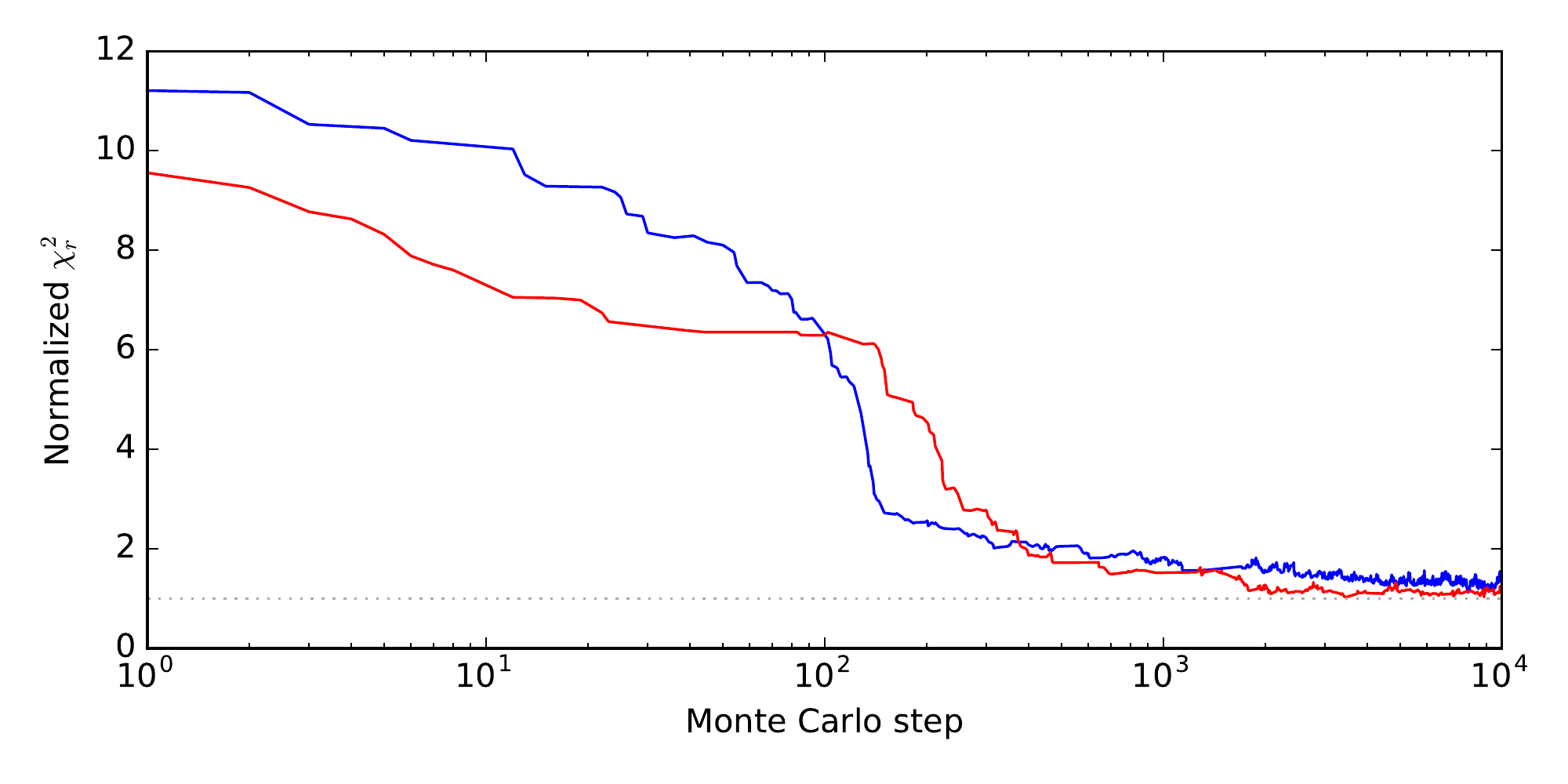}
 \caption{\label{fig:alg-converge} The evolution of the $r$-process chi-squared evaluated at successful Monte Carlo steps. The red curve indicates a run with the Monte Carlo parameters set to zero and the blue curve starts with Monte Carlo parameters varied randomly. Despite the difference in starting points, both runs eventually converge to the same solution after roughly 10,000 steps. }
\end{figure}

Because the predictions of measured masses remain relatively unchanged due to our functional form from Eq.\ (\ref{eq:masses}), we typically drop the second term of Eq.\ (\ref{eq:L}). 
In this case, the evolution of the Markov chain is only driven by the match of the network calculation to the solar abundances. 

\subsection{Convergence \& error bars}\label{sec:errorbars}
The Monte Carlo procedure outlined in the previous subsection is repeated many times until the algorithm converges. 
At the end of each step in the Metropolis algorithm, the likelihood function is calculated to determined whether or not a step is successful. 
The running average of a parameter is then computed by averaging a list of current and past values. 
If the step is successful the current value of the parameter is appended to this list. 
Otherwise, the step is not successful and the value of the parameter from the most recent past successful step is used, which may fall back to the value of the parameter from the original step. 

We take the criterion for convergence to be that the running average of all of the parameters are within their respective standard deviations. 
This definition of convergence provides a necessary condition for reaching maximum likelihood since the running average of each parameter encodes the entire evolution of the Markov chain. 
When the algorithm is near maximum likelihood, the corresponding values of the parameters are averaged over with high occurrence, thus making them more important than the starting parameters. 
Parameters that have a large influence on the results will converge to some value but will have a very small variance, while those parameters with little impact will show a larger variation with mean of the original parameter value. 

An example of the evolution of the $r$-process chi-squared, $\chi^2_{r}(m)$, for successful steps is shown in Fig.\ \ref{fig:alg-converge}. 
It takes approximately 10,000 steps for the algorithm to find the solution starting with Monte Carlo parameters set to zero (red curve) or starting from a random set of values for Monte Carlo parameters (blue curve). 
Both of these curves converge to the same set of Monte Carlo parameters, which shows that our solutions are independent of starting position. 

\begin{figure}
 \includegraphics[width=160mm]{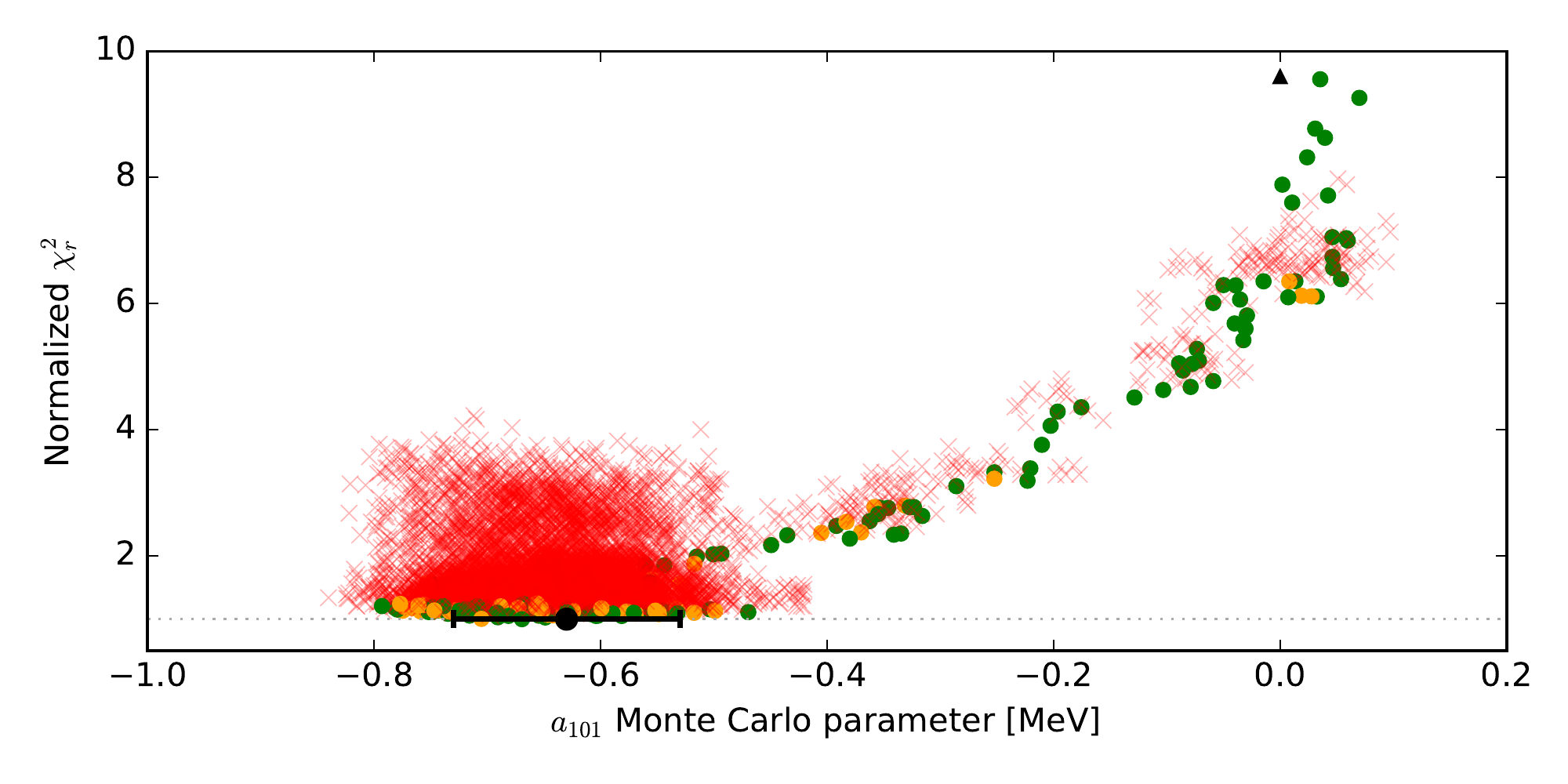}
 \caption{\label{fig:hmp} The progression of the $r$-process chi-squared as a function of the $a_{101}$ Monte Carlo parameter in the case of a very neutron-rich cold $r$-process, traj. 8. The starting point (black triangle) is $a_{101}=0$. Success steps are shown by filled green circles with failures denoted by red X's. Steps taken as successful with probability $\alpha$ from Eq.~(\ref{eq:alpha}) are shown with yellow circles. A clear trend in this variable is observed with final prediction $a_{101}=-0.63\pm 0.1$ MeV denoted by the black circle and error bars. }
\end{figure}

An example of the Markov chain evolution for a parameter in a very neutron-rich cold $r$-process is shown in Fig.\ \ref{fig:hmp}. 
The black triangle represents the starting value of zero, while successful steps are shown by green dots. 
A yellow dot represents a step counted successful with probability $\alpha$ from Eq.\ (\ref{eq:alpha}) and failure steps are denoted with a red X. 
This Markov chain produces a final prediction of $a_{101}=-0.63\pm0.1$ MeV denoted by black dot and error band. A small final error bar is seen in this Monte Carlo parameter which means it has a strong influence on the solution, as will be discussed in the results section. 

\section{Results}\label{sec:results}
With the procedure outlined in Sec.~\ref{sec:method}, we have defined a way of providing feedback to the nuclear physics by constraining our nuclear parameter space to be that which best matches the observed solar isotopic rare earth abundances. 
We are now ready to apply this framework to a number of astrophysical trajectories to reverse engineer the relevant nuclear properties important for the formation of the rare earth peak in each case. 

\begin{figure}
 \includegraphics[width=160mm]{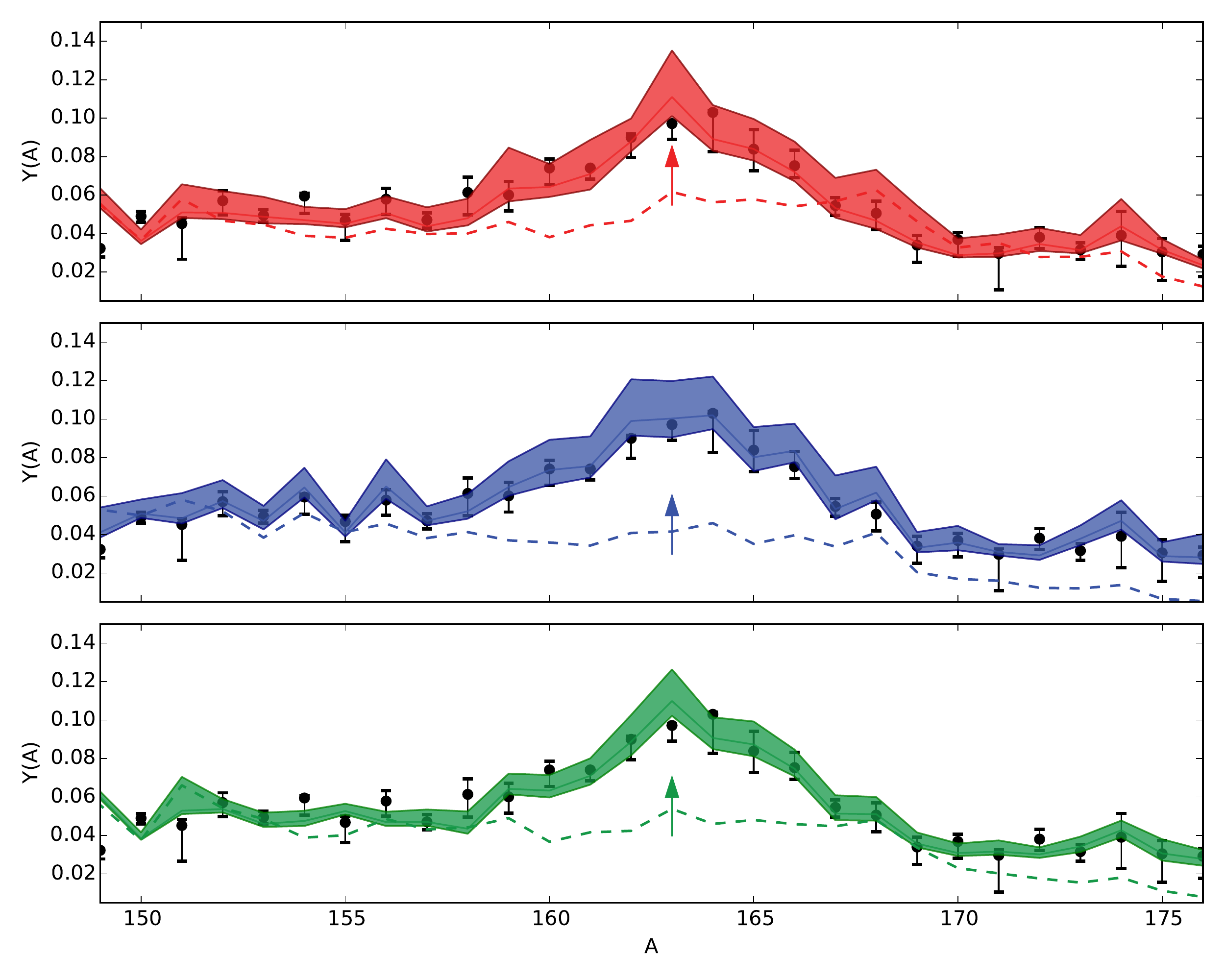}
 \caption{\label{fig:ab3i} Final abundance predictions in the rare earth region for individual trajectories: hot - traj.\ 1 (top panel), cold - traj.\ 4 (middle panel) and very neutron-rich cold - traj.\ 8 (bottom panel) Metropolis runs. The application of the reverse engineering framework starts with the baseline (dashed curves) and produces the shaded region in each case. }
\end{figure}

\subsection{Persistent rare earth feature}\label{results:persistent}
The trend in the mass surface that is responsible for the formation of the rare earth peak may be persistent, which means it spans a large range in proton number, or it may be more localized in $Z$. 
We first discuss results assuming a persistent feature with $f=40$ held fixed and the $a_N$s and $C$ allowed to vary. 

The resultant final abundances with associated error bands are shown in Fig.~\ref{fig:ab3i} for individual astrophysical trajectories. 
The error bands represent the standard deviation of previous steps as described in Sec. \ref{sec:errorbars}. 
The simulations begin with unmodified Duflo-Zuker masses, which produce the abundance patterns given by the dotted curves. 
The application of our framework produces final abundance bands that are within the solar isotopic uncertainties for each $A$ in the rare earth region. 
This shows the success of our algorithm and further indicates that our assumed abundance uncertainty of $0.1$ dex for each isotopic abundance point is a very good approximation to the real uncertainties in the rare earth region. 

The top panel of Fig.~\ref{fig:ab3i} shows a low entropy hot evolution (traj.\ 1), the middle panel shows a cold evolution (traj.\ 4) and the bottom panel shows a very neutron-rich cold evolution (traj.\ 8). 
The abundance patterns (including the baselines) are scaled to the $A\sim 150-180$ region for each set of conditions using the final simulation data from the Metropolis run. 
We find that in all cases the application of our framework successfully fills in the missing rare earth peak, and in the cold and very neutron-rich cold $r$-process conditions it also repairs the underproduction of material to the right of the peak. 

\begin{figure}
 \includegraphics[width=160mm]{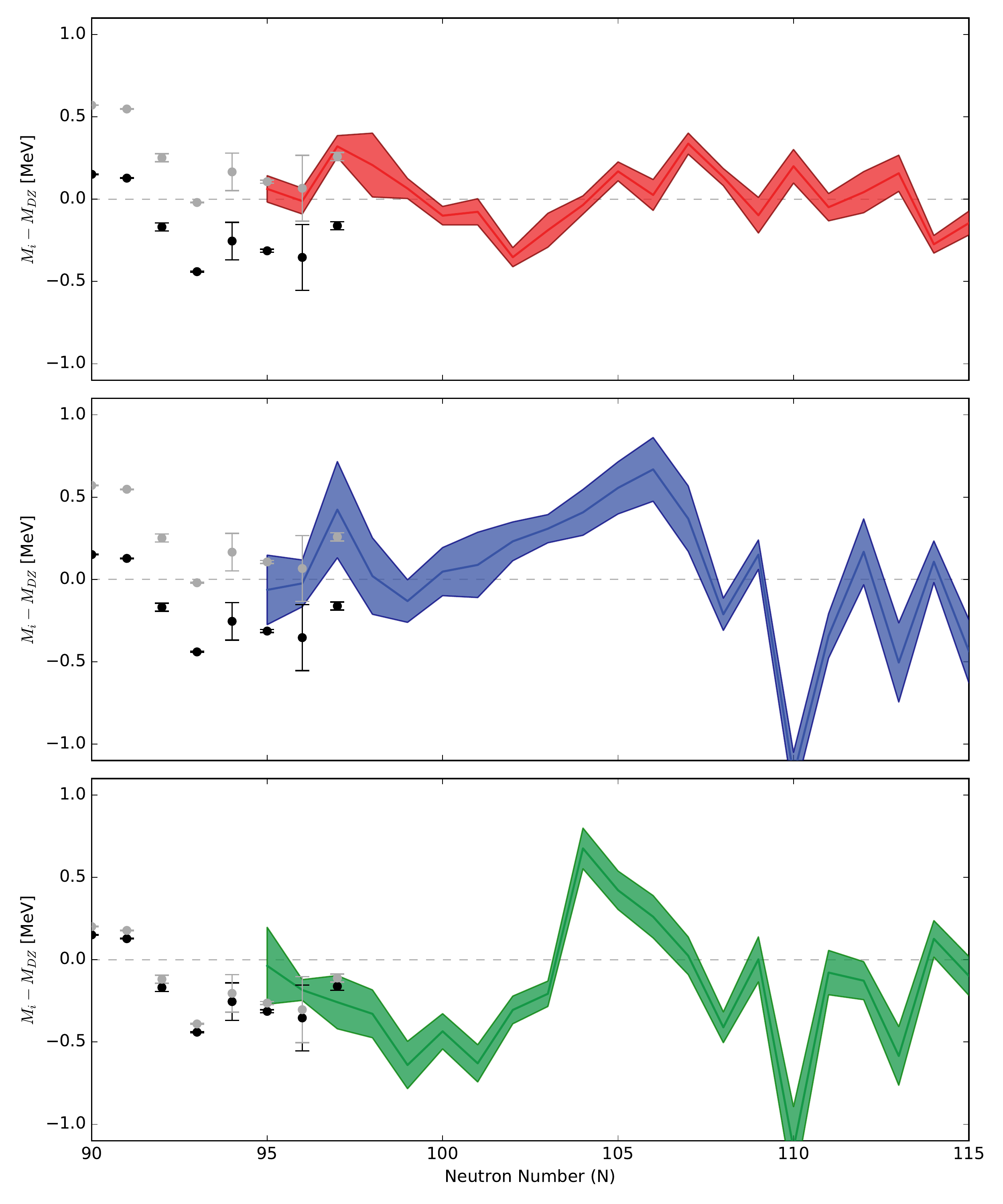}
 \caption{\label{fig:m3pi} Predicted trends in the masses of $Z=60$ neodymium isotopes for individual trajectories: hot - traj.\ 1 (top panel), cold - traj.\ 4 (middle panel) and very neutron-rich cold - traj.\ 8 (bottom panel) $r$-process conditions assuming a persistent feature ($f=40$) in proton number after application of our reverse engineering framework. The 2012 AME masses are shown in black and shifted to match the mass surface prediction in gray. }
\end{figure}

We now seek to understand the trend in the mass surface responsible for the rare earth peak production. 
The predicted trends in the masses after application of our framework are shown in Fig.~\ref{fig:m3pi} for the same individual trajectories as in Fig.~\ref{fig:ab3i}. 
Each mass surface shows a relative dip in the curve around $N\sim100$. 
The dip represents a region that has locally enhanced stability, allowing material to be hung up when the $r$-process path passes through it. 
This is the feature which is responsible for the formation of the rare earth peak in the associated panels of Fig.~\ref{fig:ab3i}. 

We stress that it is the relative, overall trends in the masses that are important for rare earth peak formation, and not the absolute values of the masses. 
Thus in Fig.~\ref{fig:m3pi} we compare our resulting mass predictions to the AME2012 measurements in two ways: the black points show the raw values while the grey points are shifted to match the predicted curves from our algorithm. 
It is clear from the comparisons between our predictions and the shifted mass data that at this time one cannot rule out any of the possible mass surfaces without more measurements to constrain the trend in the region. 

All trajectories require a dip to produce the rare earth peak, however, the trends of the mass surfaces are distinct in both the depth of the dip and its location depending on the astrophysical conditions. 
With the hot evolution, the dip is relatively shallow spanning no more than $0.8$ MeV from highest to lowest point. 
In the very neutron-rich cold evolution, the dip is stronger, spanning over $1$ MeV between highest and lowest points. 

The position of the local minimums relative to the Duflo-Zuker masses also differs as shown in Fig.~\ref{fig:m3pi}. 
For the hot evolution (top panel) the minimum is at $N=102$. 
For both cold (middle panel) and very neutron-rich cold (bottom panel), the minimum is shifted to lower $N$, consistent with an initial formation of the peak at lower mass number, $A$. 
Cold evolutions tend to have a greater availability of neutrons at late times than hot scenarios, from fission and/or from the extra $\beta$-delayed neutron emission that comes from a path very far from stability. 
Thus we find the most favored solutions tend to initially populate a rare earth peak at lower $A$, and late-time neutron captures shift the peak to the correct placement. 
The position of the minimum may be around $N=99$, or $N=101$ in these two colder scenarios. 
The inclination to favor even-$N$ in hot scenarios and odd-$N$ in cold scenarios is connected to the rare earth peak formation mechanism. 
When \nggn \ equilibrium persists for long times, such as in the hot conditions, a buildup of material occurs in even-$N$ nuclei \cite{Surman+97}. 
For colder scenarios, the path is entirely out of equilibrium and neutron capture rates are more important, thus favoring a dip at odd-$N$ nuclei \cite{Mumpower+12a, Mumpower+12c}. 

A second strong feature is noted near $N=110$ in cold and very neutron-rich cold scenarios. 
We find this feature reduces the deficiencies seen to the right of the rare earth peak that exists in the baseline model, as observed in the bottom two panels of Fig.~\ref{fig:ab3i}. 

The position in the $Z$ of maximal change in the masses from Duflo-Zuker is represented by the Monte Carlo parameter $C$ in our parameterization. 
Assuming a persistent feature, each trajectory studied shows $C$ converge to $Z=60$, or the neodymium isotopic chain. 
In our calculations, the final uncertainty for this parameter ranges from $0.1$ to $0.6$ in units of $Z$ depending on the astrophysical conditions considered.  

Figs.~\ref{fig:ab3i} and \ref{fig:m3pi} illustrate the success and power of our method. 
It would be of limited use, however, if we found distinct and dissimilar solutions for each individual astrophysical trajectory attempted. 
We find instead the exact opposite---similar mass surface trends are predicted for astrophysical trajectories within each late-time evolution characterization: hot (traj.\ 1-3), cold (traj.\ 4-6), and very neutron-rich cold (traj.\ 7-9).  

\begin{figure}
 \includegraphics[width=160mm]{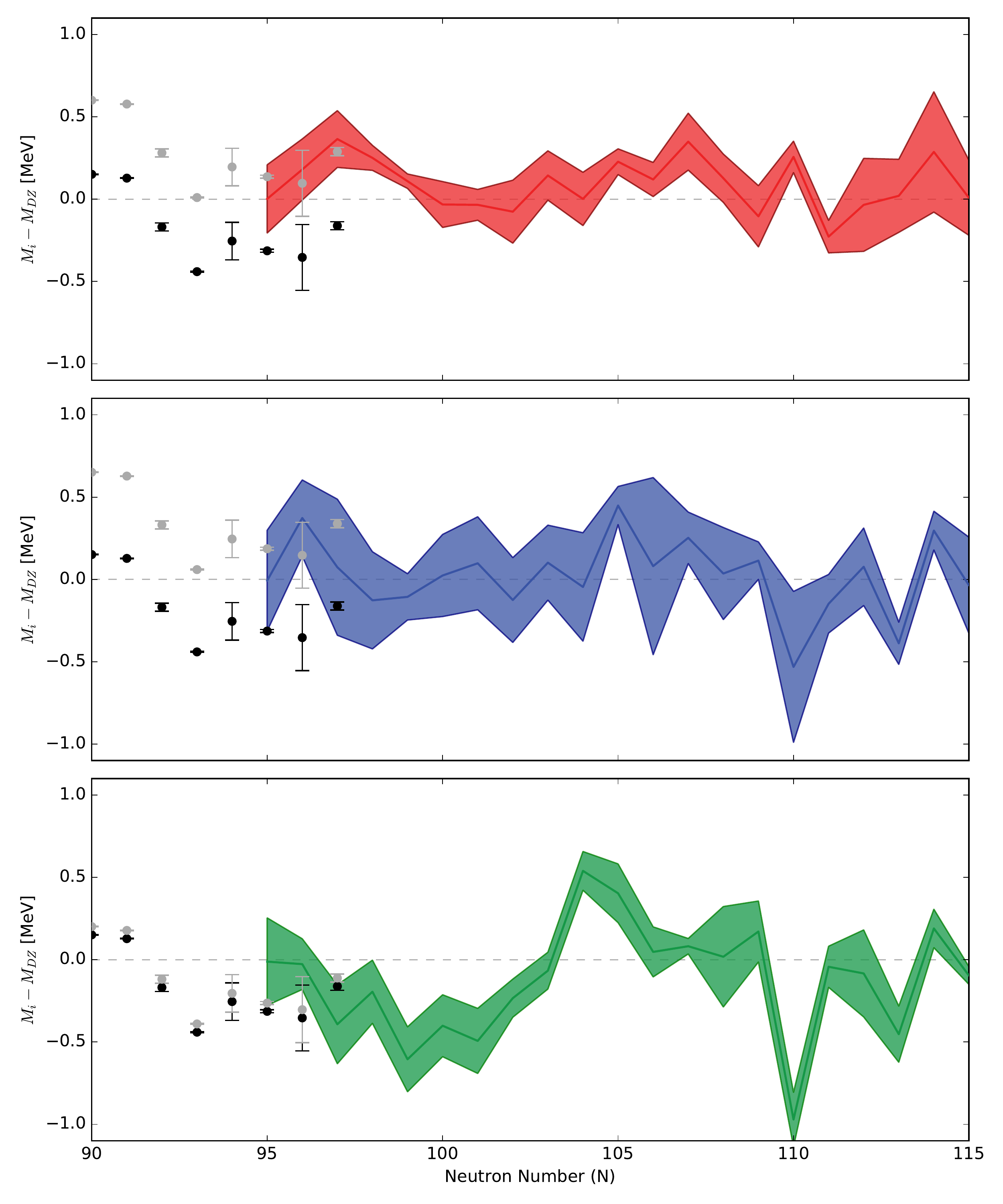}
 \caption{\label{fig:m3pc} Combined mass surfaces for hot (top panel), cold (middle panel), and very neutron-rich cold (bottom panel) $r$-process conditions assuming a persistent feature ($f=40$) in proton number. The error band in these calculations represents a spreading of the parameter space which produces the rare earth peak due to the differences in evolution of similar astrophysical conditions. }
\end{figure}

Fig.~\ref{fig:m3pc} shows the combined results of our Metropolis runs with hot trajectories 1-3, cold trajectories 4-6, and very neutron-rich cold trajectories 7-9. 
Again the predicted trends in the mass surfaces are shown along the $Z=60$ (neodymium) isotopic chain. 
The error bands for the mass surface now represent a spreading of the parameter space from the combination of the three best fit solutions of similar late-time evolutions. 
Despite the spreading of the error bands, each group of similar astrophysical conditions retains the overall trends found in the previous individual runs of Fig.~\ref{fig:m3pi}. 
Variation in similar conditions does however blur the exact location of the local minimums. 
For hot conditions the minimum may be around $N=100$, $102$, and $104$ while the position of the minimum may be around $N=97$, $99$ and $101$ in the colder scenarios. 
The observation that the overall trends remain the same, and that there are only shifts in the local minimums to even-$N$ or odd-$N$ nuclei, implies that the mechanisms for peak formation are the same for similar astrophysical conditions. 
This suggests that future measurements in this region have the potential to uncover trends in the mass surface that might point to characteristics of the $r$-process site. 

\subsection{Localized rare earth feature}\label{results:local}
There is also the possibility that the feature responsible for rare earth peak formation is more localized in proton number. 
We explore this by setting the falloff parameter to a fixed value of $f=10$ and allowing the $a_N$s and center of the strength in proton number, $C$, to vary. 

\begin{figure}
 \includegraphics[width=160mm]{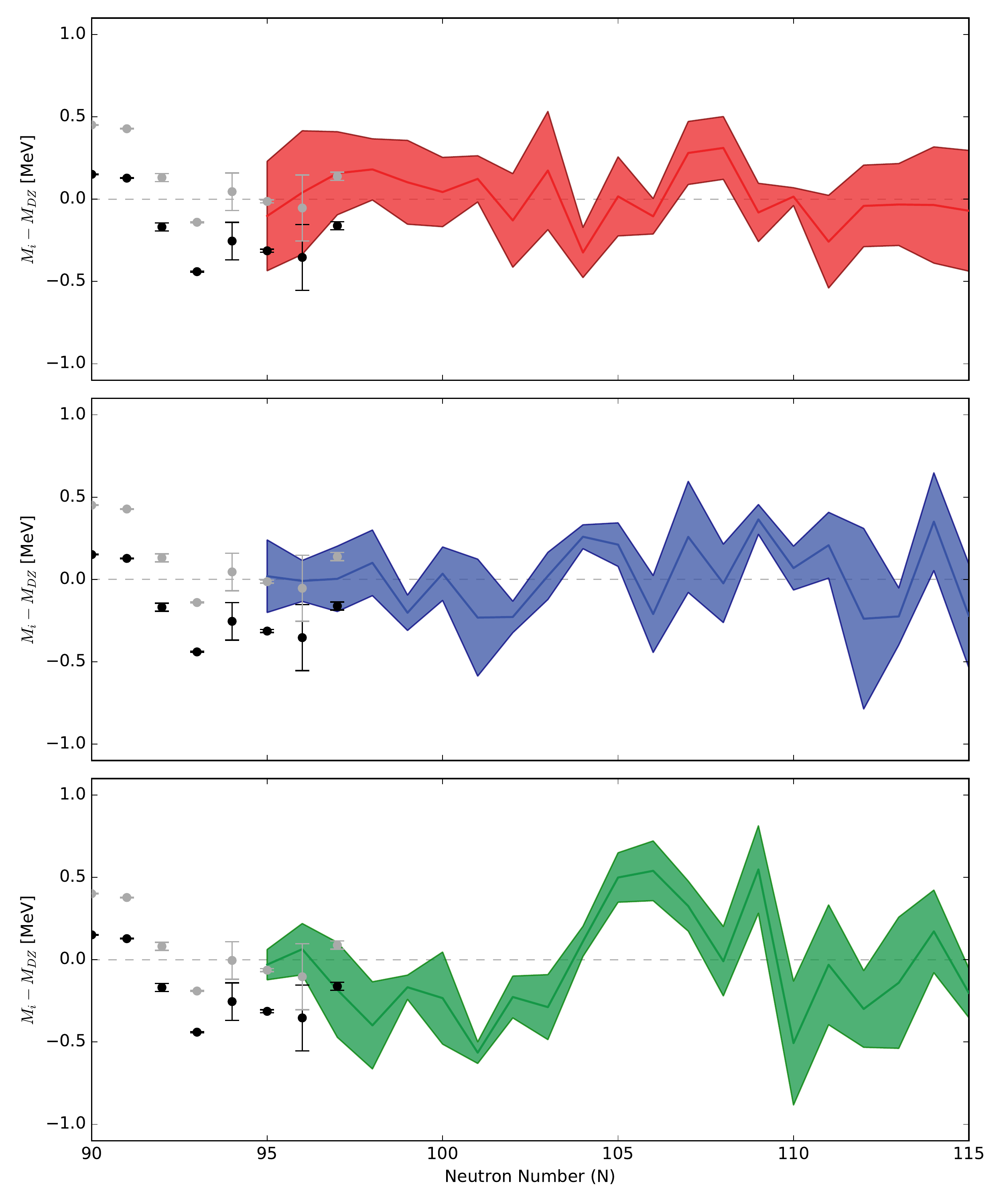}
 \caption{\label{fig:m3lc} Combined mass surfaces for hot (top panel), cold (middle panel), and very neutron-rich cold (bottom panel) assuming a localized feature ($f=10$) in atomic number. The error bands again represent a spreading of the parameter space that produces the rare earth peak due to the differences in evolution of similar astrophysical conditions. }
\end{figure}

The combined results for the more tightly localized mass surfaces predicted for each type of astrophysical conditions are shown in Fig.~\ref{fig:m3lc}. 
For the colder scenarios, the trends are similar those in Fig.~\ref{fig:m3pc}, and the $C$ parameter converges to the same value of $Z\sim 60$. 
The combination of the hot conditions, however, show less of a discernible trend. 
It is not that the rare earth peak is no longer formed, or that the algorithm does not converge. 
Rather, we find a large variation in the predicted mass surface trends and different values for $C$ among the three hot evolutions; when the three are combined the individual details of each are washed out and large error bars remain. 

These results can be understood from the different freeze-out behaviors of the trajectories. 
The spread of a broad, persistent mass feature can accommodate a range of path freeze-out positions, but a more localized feature must be tuned carefully to each individual scenario. 
In particular, the $r$-process paths at freeze-out for the hot trajectories are close to stability, with the exact locations depending sensitively on the temperature and density at neutron exhaustion. 
As a result, our algorithm finds three distinct solutions for the three hot trajectories considered here. 
The discussion of the connection between the predicted mass surface features and freeze-out dynamics continues in the next section.  

\subsection{Freeze-out dynamics and depth of rare earth dip}
As we have seen, the mechanism for rare earth peak formation couples the trends in the masses with the astrophysical conditions. 
We now discuss this in a more quantitative fashion by linking the depth of the predicted feature in the mass surface to the late-time $r$-process dynamics. 

We define the abundance-weighted shift in atomic mass number during rare earth peak formation due to neutron captures and photodissociations as
\begin{equation}
 \Delta A_\textrm{PF} = \int^{t(\frac{\tau_\beta}{\tau_{n\gamma}}\sim1)}_{t(R\sim1)} \lambda_{n\gamma} - \lambda_{\gamma n} \textrm{d}t
\end{equation}
where $\lambda_{n\gamma}$ and $\lambda_{\gamma n}$ are the abundance-weighted average neutron capture and photodissociation rates, respectively, in units of sec$^{-1}$, and the integration is performed from the time at which the neutron-to-seed ratio, $R$, is unity until the abundance-weighted timescales for neutron capture and $\beta$-decay are roughly equal. 
A definition of abundance-weighted timescales and their inverses, the $\lambda$'s, can be found in Ref.~\cite{Mumpower+12b}. 
We define the late-time shift in atomic mass number due to neutron captures as
\begin{equation}
 \Delta A_\textrm{LT} = \int_{t(\frac{\tau_\beta}{\tau_{n\gamma}}\sim1)}^{t(\textrm{end})} \lambda_{n\gamma} \textrm{d}t
\end{equation}
where the range of time is between when the abundance-weighted timescales for neutron capture and $\beta$-decay are equal and the end of the simulation. 
The second term in the integrand of Eq.\ (9) does not appear in Eq.\ (10) since it is negligible for $t>t(\frac{\tau_\beta}{\tau_{n\gamma}}\sim1)$. 
The rate net neutron capture during rare earth peak formation can be approximated by taking the ratio of $\Delta A_\textrm{PF}$ to the difference in time, $\Delta t$, associated with the integration range.

\begin{center}
\begin{longtable}{@{\extracolsep{\fill}} c|c|c|c|c|c|c}
\caption{\label{table2} Information on rare earth peak formation for each trajectory studied in this work, from the reverse-engineering runs that assume a persistent ($f=40$) feature as described in Sec.\ 3.1. Definitions of each column are found in the text. }
\setlength{\tabcolsep}{2cm}\\
Trajectory & Type & $\Delta A_\textrm{LT}$ & $\Delta A_\textrm{PF}$ & $\Delta$t & $\Delta A_\textrm{PF}$/$\Delta t$ & Dip depth [MeV]\\
\hline\hline
\endhead
traj.\ 1 & hot              & 0.4 & 3.8 & 0.23 & 16.5 & 0.68 \\
traj.\ 2 & hot              & 0.5 & 3.1 & 0.18 & 17.3 & 0.72 \\
traj.\ 3 & hot              & 0.6 & 2.7 & 0.15 & 18.0 & 0.74 \\
traj.\ 4 & cold             & 1.9 & 0.9 & 0.04 & 22.5 & 1.13 \\
traj.\ 5 & cold             & 1.7 & 1.4 & 0.06 & 23.3 & 0.98 \\
traj.\ 6 & cold             & 2.0 & 0.9 & 0.04 & 23.1 & 1.03 \\
traj.\ 7 & very n-rich cold & 1.9 & 1.3 & 0.05 & 26.0 & 1.16 \\
traj.\ 8 & very n-rich cold & 2.2 & 1.4 & 0.05 & 28.0 & 1.32 \\
traj.\ 9 & very n-rich cold & 2.1 & 1.7 & 0.06 & 28.3 & 1.34
\end{longtable}
\end{center}

In Table \ref{table2} we compare the freeze-out quantities defined above to the reverse-engineering results from Sec.~\ref{results:persistent}. 
In that section we found that the results for hot trajectories have a dip in the mass surface centered at higher neutron number \textcolor{blue}{than} the results for the colder trajectories \textcolor{blue}{do}. 
The late-time shift in atomic mass due to neutron captures, $\Delta A_\textrm{LT}$, shows a clear difference between the warmer and colder evolutions. 
The distinct rare earth peak formation mechanisms identified for the hot and cold trajectories by our reverse-engineering studies are attributed to this difference. 
In hot evolutions there is little late-time bulk transfer of material in $A$, thus the peak forms in the right spot. 
In contrast, the greater availability of neutrons late in freeze-out in the colder evolutions favors the peak to form off-center, to the left in atomic mass, with the final placement achieved by late-time neutron captures. 

We find that a faster movement of material through the rare earth region during peak formation requires a larger dip in the mass surface, as seen comparing the last two columns of Table \ref{table2}. 
This near linear relationship suggests that future mass measurements which find a trend may be able to shed light on how quickly the $r$-process path moves through the region of the $NZ$-plane where the peak is formed. 

\subsection{Systematic uncertainties}\label{sec:uncert}
Calculations of the $r$ process have many theoretical uncertainties which may impact the application of the reverse engineering framework. 
We now cover several possibilities and discuss the impact on our conclusions. 

\begin{figure}
 \includegraphics[width=140mm]{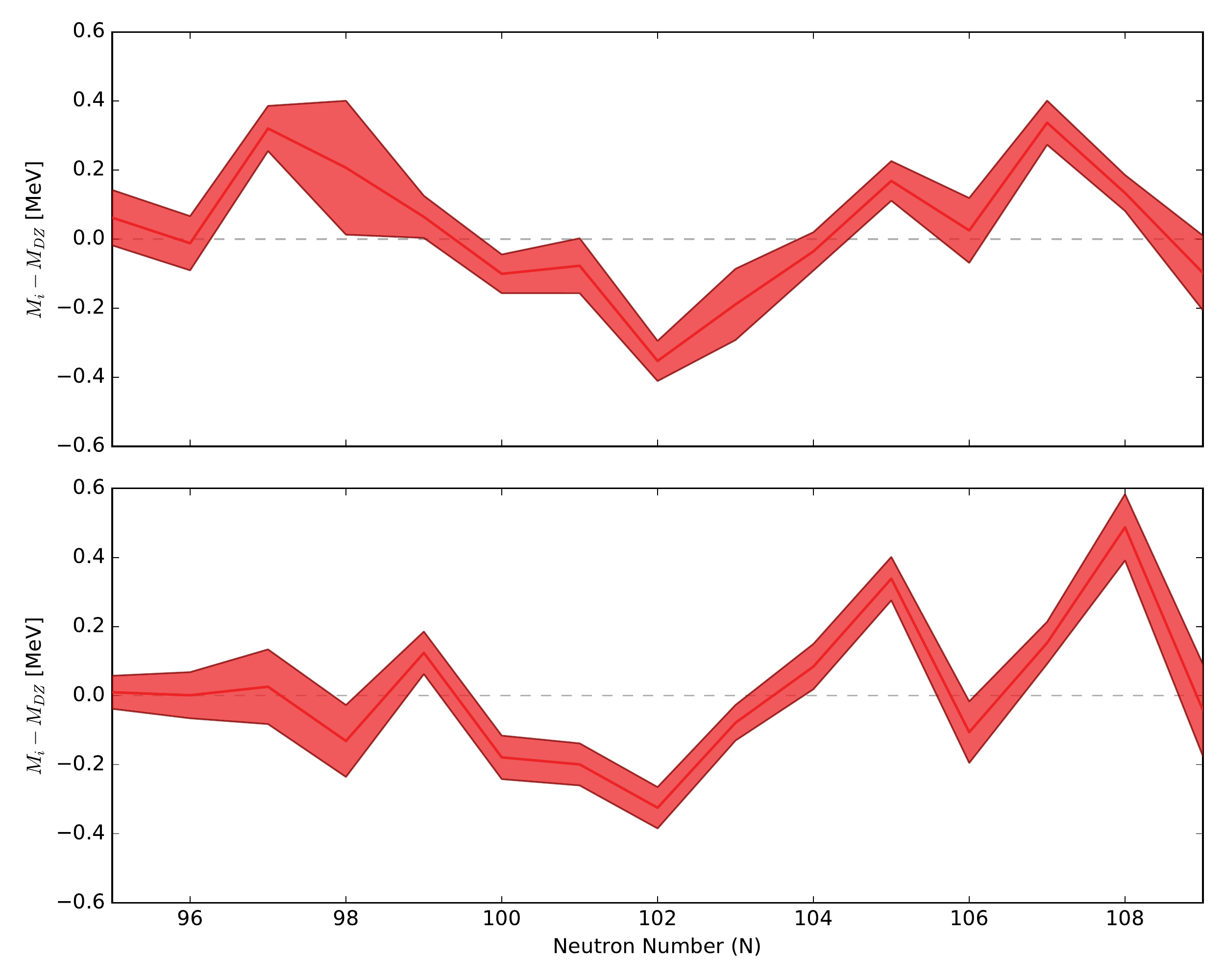}
 \caption{\label{fig:ncap_masses} The predicted trends in the mass surface remain relatively unchanged from the use of CoH neutron capture rates (top panel) assuming different neutron capture rates (Nonsmoker - bottom) in a hot, low entropy $r$ process (traj.\ 1).  }
\end{figure}

Model predictions of neutron capture rates generally range between a factor of 3 in the rare earth region using similar model inputs of LD, $\gamma$SF and OMP, as shown in Fig.~\ref{fig:tncap}. 
To test whether this variation in rates has an impact on our results, we implement the results of a second statistical model code, NONSMOKER \cite{Rauscher+00}. 
The available compilation of NONSMOKER rates was calculated using FRDM1995 masses. 
Thus, we first perform the same fitting procedure as in Sec.~\ref{method:ncap} to generate a set of temperature-dependent parameters $a(N,T)$, $b(N,T)$, $c(N,T)$ from the compiled NONSMOKER rates and the FRDM1995 masses. 
The reverse engineering framework is then used to find the mass surface responsible for rare earth peak formation, using the NONSMOKER capture rate parameters $a(N,T)$, $b(N,T)$, and $c(N,T)$. 

We compare the mass surface with NONSMOKER rates (bottom panel) to the results of the baseline CoH rates (top panel) in Fig.~\ref{fig:ncap_masses} for the case of the hot, low entropy $r$ process. 
We find that both the minimum position of the dip and the overall trend in neutron number of the mass surface remains relatively unchanged with the change of neutron capture rate datasets. 
This suggests that our conclusions are fairly robust in terms of reasonable variations in neutron capture rates. 

It may be argued that still larger systematic uncertainties plague predictions of neutron capture rates in the rare earth region from missing physics in the model inputs. 
For example, an enhancement of the $\gamma$SF with missing M1 strength from Low Energy MAgnetic Radiation or `LEMAR' was suggested recently by Refs.~\cite{Schwengner+13, Frauendorf+14}. 
Preliminary calculations suggested that including the low-lying M1 strength could result in rates larger by a factor of $5$ to $10$, which could impact the rare earth region \cite{Frauendorf+15}, while more recent calculations \cite{Kawano+16b} found a smaller influence on the rates and a minimal impact on the final $r$-process abundances. 
Ref.~\cite{Kawano+16b} also found that switching between the Koning OMP and the deformed Kunieda OMP made very little difference (on the order of a factor of 2 or less) in the predictions of neutron capture rates in the rare earth region. 
The impact of LD calculations on rare earth abundances has yet to be studied and will be the subject of future work. 

Modern predictions of rare earth $\beta$-decay rates are in fairly good agreement, showing roughly a factor of $2$ deviation between model calculations at high $Q_{\beta}$ \cite{Moller+03, Marketin+16, Shafer+16}. 
We explored the impact of this factor of two rate deviation on our reverse-engineering studies in Ref.~\cite{Mumpower+nic16}. 
We reran a selected set of the studies from Sec.\ 3.1 with rare earth $\beta$-decay rates everywhere either increased or decreased by a factor of two. 
We found that the height of the rare earth peak and the extent of the feature in the mass surface were both changed. 
If $C$ was held constant and the $\beta$-decay rates were increased, the dip in the rare earth mass surface would become larger to compensate for the change. 
For a slowdown in $\beta$-decay rates, the dip in the rare earth mass surface became more shallow. 
In either case of speeding up or slowing down rare earth $\beta$-decay rates, the mechanism for peak production remained the same in both hot and cold environments. 

\begin{figure}
 \begin{center}
  \includegraphics[width=\textwidth]{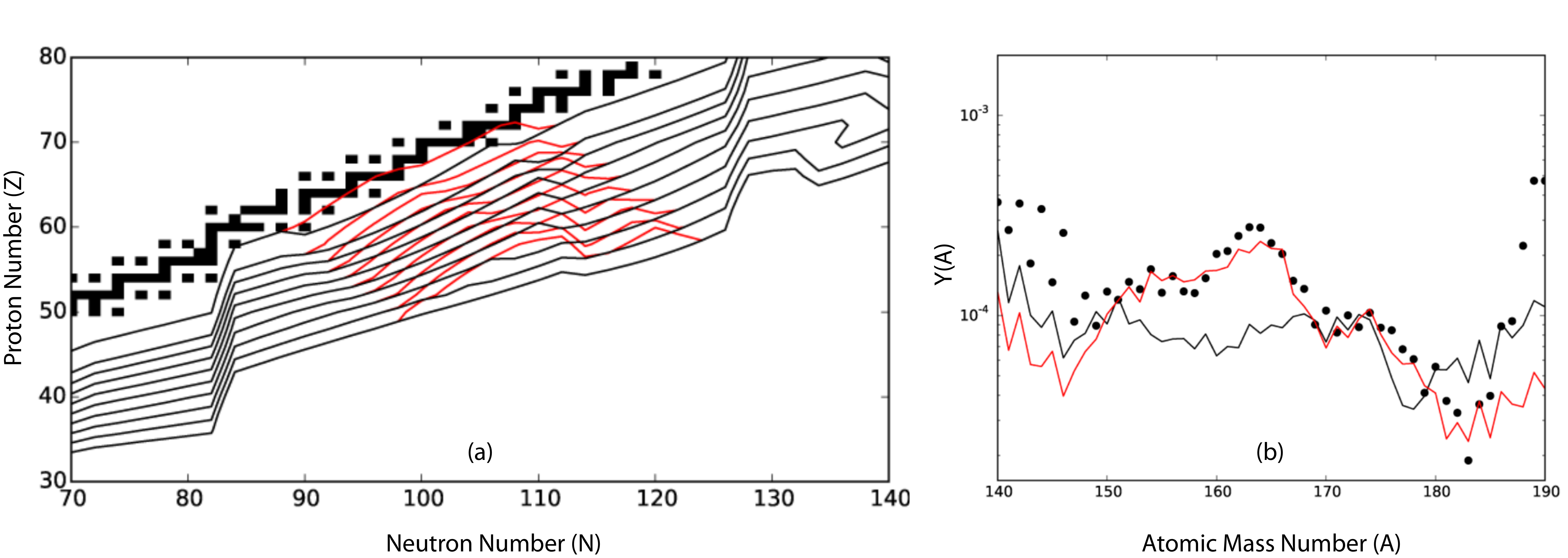}
  \caption{\label{fig:dz} One neutron separation energies from DZ (Left panel) and final rare earth peak (right panel) shown in black. The same data is again shown in red when using the algorithm to attempt to reproduce the observed abundances. In this case, the parameter set does not simultaneously match the known masses and can be ruled out. }
 \end{center}
\end{figure}

Another possible starting point for our reverse engineering framework is to use the parameters of the Duflo-Zuker mass model as our Monte Carlo parameters. 
Using feedback only from the rare earth abundances, e.g., only Eq.\ (\ref{eq:chi2r}) is used in the calculation of the likelihood function, we find the mass surface (red lines) shown in the left panel of Fig.~\ref{fig:dz}. 
In this case the Monte Carlo parameters successfully produce the rare earth peak (red curve in the right panel), however, it does so at the expense of the match to measured masses. 
If we constrain both the measured masses and the rare earth abundances by including both Eqs.\ (\ref{eq:chi2r} \& \ref{eq:chi2m}) in the likelihood function, we can find no combination of parameters that produces the rare earth peak. 
This result shows that additional parameters are needed to explain the formation of the rare earth peak using this mass model. 

\section{Summary}\label{sec:summary}
While there are large uncertainties in the inputs to $r$-process nucleosynthesis, the output---the pattern of solar $r$-process residuals---is relatively well known. 
This opens up the $r$ process to treatment as an inverse problem. 
Here we have developed a Monte Carlo framework to reverse engineer unknown nuclear properties using a quantitative match to the solar isotopic pattern, starting from a range of different astrophysical conditions. 
Ultimately, we aim to correlate engineered nuclear structure features to characteristics of possible $r$-process environments, such that future experiments can search for these features and thus help to constrain the $r$-process site. 

In this work, we have applied our reverse-engineering framework to the neutron-rich rare earth region in an attempt to understand the mass trends responsible for the formation of the rare earth peak. 
Our procedure starts with Duflo-Zuker masses, which are featureless in the rare earth region and produce flat abundance predictions, and finds solutions with mass modifications to Duflo-Zuker that reproduce the rare earth peak to within the solar isotopic pattern uncertainties. 
We look for two types of solutions: those that result in a persistent feature in the mass surface that spans a large range in proton number $Z$, and those which produce a feature more localized in $Z$. 
In both cases, the trends found in the mass surface responsible for rare earth peak production depend on the adopted astrophysical conditions. 

When a persistent feature is assumed, we find traditional, hot $r$-process trajectories that go through a long duration \nggn \ equilibrium require trends in masses near $Z=60$ neodymium isotopes that have local minimums at even-$N$ nuclei near $N\sim100$ and span a change of no more than $0.8$ MeV. 
Colder $r$-process trajectories that have a short duration \nggn \ equilibrium are found to require trends in the mass surface that have local minimums at odd-$N$ and span a change of over $1$ MeV. 
We find that the depth of the feature in the masses near $N\sim100$ is directly related to how far the $r$-process path proceeds towards the neutron dripline and how fast it moves back to stability. 
In all cases, the trends in the predicted masses are extended in neutron number and are not the abrupt changes that might be expected, e.g., from a subshell closure. 
Nuclear deformation is a possible source of these smooth trends. 

When we look for a localized feature, we find solutions that depend more sensitively on the details of the astrophysical conditions. 
This is most pronounced in the case of hot trajectories where we find a larger deviation between the resultant mass surfaces. 

Our results suggest that a wealth of information can be obtained from new measurements in the rare earth region. 
If a sizable region of enhanced stability is found, its characteristics could point to the nature of the $r$-process site: hot, cold, or very neutron-rich cold. 
More detailed information about freeze-out conditions could potentially be extracted from the location and depth of a small, localized region of enhanced stability. 
The absence of any significant feature would disfavor the dynamical method of rare earth peak formation. 
This would point instead to a rare earth peak composed of fission fragments, which would argue for neutron star mergers as the main $r$-process astrophysical site. 
It would also be possible to use this method to consider partial fission / partial dynamical solutions for any given prediction of fission rates and daughter distributions. 

The past few years has seen a dramatic increase in the quantity and quality of experimental data for neutron-rich nuclei important for the $r$ process, e.g., \cite{Hakala+12, Madurga+12, VanSchelt+13, Kurtukian+14, Caballero-Folch+14, Spyrou+14, Sun+15, Atanasov+15, Klawitter+15, Lascar+15, Lorusso+15, Cizewski+15, Mazzocchi+15, Jones+15, Dunlop+16, Alshudifat+16, Liddick+16, Domingo+16, Miernik+16, Caballero-Folch+16, Hirsh+16, Wu+16}.
Future measurement campaigns at current and planned experimental facilities such as the Facility for Rare Isotope Beams, will offer an unprecedented access to the production of short-lived isotopes \cite{Horowitz+16r}. 
Our study has pinpointed nuclei in the rare earth region which have a substantial impact on the formation of the rare earth peak. 
A combined theoretical and experimental effort will help to distinguish between astrophysical conditions, thus providing an avenue for moving forward with the solution of the site(s) of the $r$ process. 

\section*{Acknowledgements}
This work was supported in part by the National Science Foundation through grant number PHY1554876 (AWS) and the Joint Institute for Nuclear Astrophysics grant numbers PHY0822648 and PHY1419765 (MM), and the U.S. Department of Energy under grant numbers DE-SC0013039 (RS) and DE-FG02-02ER41216 (GCM). 
A portion of this work was also carried out under the auspices of the National Nuclear Security Administration of the U.S. Department of Energy at Los Alamos National Laboratory under Contract No. DE-AC52-06NA25396 (MM). 

\section*{References}
\bibliographystyle{unsrt}
\bibliography{refs}

\end{document}